\pdfoutput=1

\documentclass[%
 reprint,
superscriptaddress,
 amsmath,amssymb,
 aps,
]{revtex4-2}

\usepackage{graphicx}
\usepackage{dcolumn}
\usepackage{bm}

\usepackage{xcolor}

\begin{document}


\title{Signatures of heterogeneity in the statistical structure of target state aligned ensembles}

\author{Nicolas Lenner}
 \altaffiliation[Currently at ]{Simons Center for Systems Biology, School of Natural Sciences, Institute for Advanced Study, Princeton, New Jersey, USA.}
  \email{Lenner@ias.edu}
 \affiliation{Max Planck Institute for Dynamics and Self-Organization, Göttingen, Germany}

\author{Matthias Häring}
\affiliation{Max Planck Institute for Dynamics and Self-Organization, Göttingen, Germany}
\affiliation{Göttingen Campus Institute for Dynamics of Biological Networks, University of Göttingen, Göttingen, Germany}

\author{Stephan Eule}%
\affiliation{Max Planck Institute for Dynamics and Self-Organization, Göttingen, Germany}
\affiliation{German Primate Center—Leibniz Institute for Primate Research, Goettingen, Germany}

\author{Jörg Großhans}
\affiliation{Göttingen Campus Institute for Dynamics of Biological Networks, University of Göttingen, Göttingen, Germany}
\affiliation{Department of Biology, Philipps University Marburg, Marburg, Germany}

\author{Fred Wolf}
 \email{Fred.Wolf@ds.mpg.de}
\affiliation{Max Planck Institute for Dynamics and Self-Organization, Göttingen, Germany}
\affiliation{Göttingen Campus Institute for Dynamics of Biological Networks, University of Göttingen, Göttingen, Germany}
\affiliation{Max Planck Institute for Multidisciplinary Sciences, Göttingen, Germany}
\affiliation{Institute for the Dynamics of Complex Systems, University of Göttingen, Göttingen, Germany}
\affiliation{Center for Biostructural Imaging of Neurodegeneration, Göttingen, Germany}
\affiliation{Bernstein Center for Computational Neuroscience Göttingen, Göttingen, Germany}


\begin{abstract}
Finite time convergence to functionally important target states is a key component of many biological processes. We previously found that the terminal approach phase of such dynamics exhibits universal types of stochastic dynamics that differ qualitatively between noise-dominated and force-dominated regimes of the approach dynamics. While for the noise-dominated regime the approach dynamics is uninformative about the underlying force law, in the force-dominated regime it enables the accurate inference of the underlying dynamics. Biological systems often exhibit substantial parameter heterogeneity, for instance 
through copy number fluctuations of key molecules or variability in modulating factors. Here, we extend our theory of target state aligned (TSA) stochastic dynamics to investigate the impact of parameter heterogeneity in the underlying stochastic dynamics. We examine the approach to target states for a wide range of dynamical laws and additive as well as multiplicative noise. We find that the distinct regimes of noise-dominated and force-dominated dynamics strongly differ in their sensitivity to parameter heterogeneity. In the noise-dominated regime, TSA ensembles are insensitive to parameter heterogeneity in the force law, but sensitive to sample to sample heterogeneity in the diffusion constant. 
For force-dominated dynamics, both parameter heterogeneity in the force law
and diffusion constant change the behaviour of the non-stationary statistics and in particular the two-time-covariance functions. In this regime, TSA ensembles provide a sensitive readout of parameter heterogeneity. Under 
natural conditions, parameter heterogeneity in many biological systems cannot be experimentally controlled or eliminated. Our results provide a systematic theoretical foundation for the analysis of target state directed dynamics in a large class of systems with substantial heterogeneity.
\end{abstract}

\maketitle



\section{Introduction}

Biological systems often comprise a multitude of interacting components, covering scales from individual molecules, to protein complexes to cells, membranes and whole tissues \cite{milo2015cell}. At the system level, their dynamics can typically be summarized by one or a few collective dynamical variables \cite{noe2017collective,stephens2008dimensionality}. Directional dynamics in living systems (i) guide the system towards functionally important target states \cite{sha2003hysteresis,pomerening2003building,coudreuse2010driving,rata2018two,schwarz2018precise,domingo2011switches,nachman2007dissecting,pardee1974restriction,ahrends2014controlling,maamar2007noise,xiong2003positive,losick2008stochasticity,balazsi2011cellular,hanes1996neural,hanks2015distinct,ratcliff2016diffusion,ratcliff2008diffusion,brunton2013rats,hanks2015distinct,churchland2011variance,roitman2002response}; (ii) are intrinsically stochastic not only due to thermal fluctuations but also the effective impact of ongoing active processes \cite{nachman2007dissecting,balazsi2011cellular,hanes1996neural}; (iii)  exhibit sample to sample variability in the details of the machinery \cite{altschuler2010cellular,komin2017address,gough2017biologically}.

We recently addressed the first two aspects of directionality (i) and stochasticy (ii) within an inference framework of target state aligned (TSA) dynamical ensembles \cite{lenner2023reverse1,lenner2023reverse2}.
For this class of directional effective dynamics, we proposed to analyze the ensemble in its natural frame of reference, i.e. to align all sample paths to the target state and to infer the dynamics in reverse time. The target state then becomes the initial condition of the newly formed reverse-time ensemble. 

We also examined the impact of state-dependent (multiplicative) noise \cite{lenner2023reverse2}. State-dependent noise is a common phenomenon, often reflecting intrinsically state dependent dynamics such as reaction kinetics \cite{gillespie2000chemical} or resulting from the projection of high dimensional dynamics onto a low dimensional representation \cite{berezhkovskii2011time}.

\begin{figure}[ht]
\centerline{\includegraphics[width=\linewidth]
{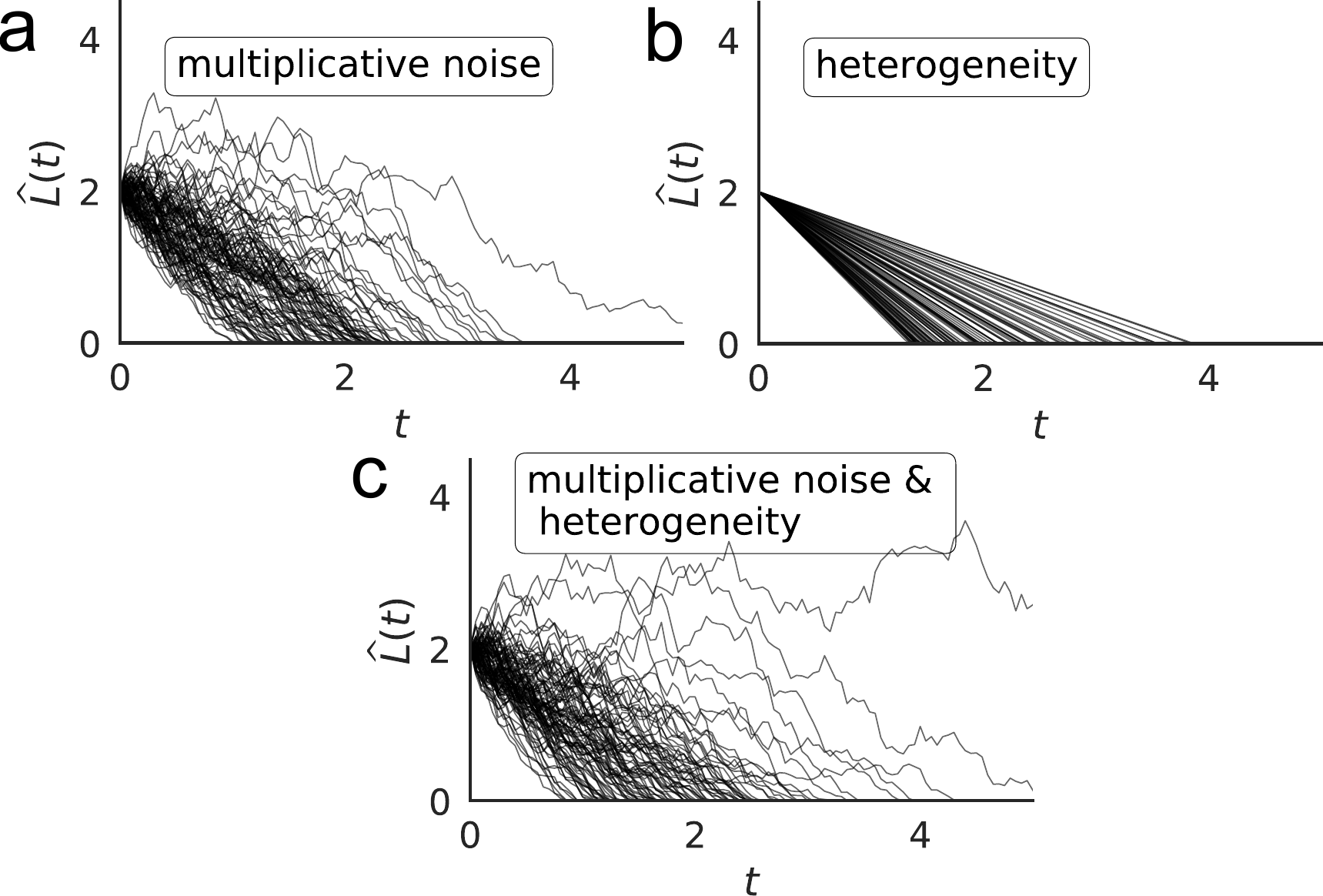}}
\caption{
\textbf{Joint variability due to heterogeneity and intrinsic stochasticity is a hallmark of biology yet hard to distinguish experimentally.} 
We show three exemplary realizations of target state directed dynamics: \textbf{(a)} shows an example for homogeneous dynamics, where all variability is due to the intrinsic stochasticity of the dynamics - here represented as multiplicative noise. \textbf{(b)} depicts the same process without noise but with sample to sample variability in the force law. This case symbolizes pure heterogeneity. Finally, \textbf{(c)} shows a superposition of both the heterogeneous and stochastic dynamics. Without careful application of statistical methods the two components can not be disentangled.
}\label{MultForceDriveTSAx3_test}
\end{figure}

In addition to intrinsic noise, biological samples typically show sample to sample heterogeneity, a feature that needs to be thoroughly understood for the application of TSA concepts \cite{altschuler2010cellular,gough2017biologically}. 
Across many samples of the same process, the general form of the dynamics may be identical, but the effective parameters that control the overall speed or noise intensity may vary from sample to sample \cite{swain2002intrinsic}, see Fig.~\ref{MultForceDriveTSAx3_test}. Such parameter heterogeneity can significantly alter the observed ensemble statistics of the dynamics compared to model predictions based on homogeneous dynamics.

 Here we generalize our treatment of TSA ensembles in reverse time to heterogeneous TSA ensembles with state dependent noise statistics. We briefly recapitulate the theory of target state alignment and key results for homogeneous dynamics  \cite{lenner2023reverse1,lenner2023reverse2}. We then extend our theory to heterogeneous dynamics and demonstrate the systematic changes which heterogeneity introduces into the observed TSA ensemble statistics. Finally, we discuss the functional form of the observed ensemble statistics in order to provide an intuition about TSA ensemble statistics as they occur in real biological systems.

\section{Theory of TSA dynamics}
In biological systems the dynamics of a generalized coordinate $\widehat{L}$ are often well approximated by a Langevin equation
\begin{align}
\label{eq:fwd_langevin}
    d\widehat{L} = f(\widehat{L}) dt + \sqrt{D(\widehat{L})} \; dW_t
\ .
\end{align}
Here, $f(\widehat{L})$ summarizes the deterministic force contribution, $D(\widehat{L})$ is the state dependent diffusion ``constant'' and $dW_t$ denotes the Wiener increment of a zero mean delta correlated $\langle dW_t, dW_{t'} \rangle = \delta(t-t')$ white noise contribution. Drawing from a distribution of initial values $P^\mathrm{in}(\widehat{L}_\mathrm{in})$, Eq.~\eqref{eq:fwd_langevin} can be used to generate sample paths that approximate the dynamics of biological processes. Throughout this text we use the Ito interpretation of the Langevin equation \cite{van1981ito}.

For the here relevant case of biological dynamics which converge towards functionally important target states, initial conditions are typically ill defined. Only close to the target state a well defined dynamical law seems to emerge from a background of other biological processes. To analyze such dynamics we proposed in Lenner et.~al. \cite{lenner2023reverse1} to align all trajectories to the target state and analyze the dynamics with respect to their time to completion $\tau=T_i-t$, where $T_i$ is the lifetime of the $i$-th samplepath. This new ``reverse-time'' $\tau$ is zero at the target state and grows as we move away from the target state. We denote time and position in such an ensemble as $(L,\tau)$. The mathematically exact mapping of the original forward dynamics to such a target-state aligned (TSA) reverse-time ensemble $R(L,\tau)$ is given in Lenner et.~al. \cite{lenner2023reverse1} and \cite{lenner2023reverse2}. 

Typically, we are only interested in the dynamics close to the target state measured in reverse time $\tau$. 
We showed in \cite{lenner2023reverse1,lenner2023reverse2}, that a reverse time Langevin equation of the form
\begin{align}
\label{tr_lv_tsa}
  dL(\tau) =
\left(
      f(L) 
      +
      f^\mathcal{F}(L)
 \right)
       d\tau
      +
      \sqrt{D(L)} \ d W_\tau
\end{align}
with the free energy force
\begin{align}
\label{freeEforce_multN_closeToCompl}
f^\mathcal{F}(L)
=
       D(L) 
      \  
      \frac{\partial}{\partial 
L}
 \log\left(
  \int_{L_\mathrm{ts}}^{L} dL' \;
   e^{-
  \int^{L'}
 \frac{2 f(L'')}{D(L'')} dL''
 } 
 \right)
 \ ,
\end{align}
describes the dynamics close to completion, that is the target state, well. For infinitely far separated initial and final states the expression is exact. Due to its form, we call Eq.~\eqref{freeEforce_multN_closeToCompl} a free energy force. It ensures that both the deterministic components of the forward dynamics and their entropy production with forward time are simultaneously reverted.

To form a better understanding of the general behavior of TSA dynamics close to target states, we expand both the force and the diffusion term in lowest order powerlaws around a target state $L_\mathrm{ts}=0$. Concretely, we study the force 
\begin{align}
 f(L) = -\gamma L^\alpha
\end{align}
and diffusion term
\begin{align}
 D(L)=D L^\beta
 \ ,
\end{align}
with $\gamma , D>0$ and $\alpha , \beta \in \mathbb{R}$. We show in \cite{lenner2023reverse2}, that dynamics with jointly $\alpha \ge 1$ and $\beta \ge 2$ cannot reach the target state at $L_\mathrm{ts}=0$ in finite time. Dynamics with any other combination of $\alpha$ and $\beta$ will eventually reach this target state.

In \cite{lenner2023reverse2} we derived a unique analytic expression which captures the alignment and time reversal of all power law dynamics which reach the target state in finite time. For $\alpha -\beta \neq -1$ we found
\begin{align}
  \label{sup_freeEforce_fwd_powerlaw_mult}
&f^\mathcal{F}(L)
=
\\
\notag
&=
L^\beta
 \frac{D(\alpha-\beta+1) \left(-\frac{2 \gamma }{D(\alpha-\beta+1)}\right)^{\frac{1}{\alpha-\beta+1}} e^{\frac{2 \gamma  L^{\alpha-\beta+1}}{D(\alpha-\beta+1)}}}{\Theta(\alpha-\beta+1)  \Gamma \left(\frac{1}{\alpha-\beta+1}\right)-\Gamma
   \left(\frac{1}{\alpha-\beta+1},-\frac{2 L^{\alpha-\beta+1} \gamma }{D(\alpha-\beta+1)}\right)}
   \ .
\end{align}
Here, $\Theta(n)$ denotes the Heaviside step-function, $\Gamma(z)$ the gamma-function and $\Gamma(n,z)$ the upper incomplete gamma-function. The connecting case at $\alpha-\beta=-1$ is given as 
$f^\mathcal{F}(L)=(2 \gamma + D) L^\alpha$.

To gain further insights into the dynamics in the vicinity of the target state, we expand the free energy force Eq.~\eqref{sup_freeEforce_fwd_powerlaw_mult} for small $L$ (see \cite{lenner2023reverse1,lenner2023reverse2}). Dependent on the relation of the power law exponents, we find two regimes for the respective TSA reverse-time Langevin equation. The case $\alpha \ge \beta$:
 \begin{align}
\label{sup_tr_lv_tsa_sn_multN_alpha_ge_beta_simp}
  dL(\tau) =
&\left(
      D L^{\beta-1}
      -
      \gamma \frac{\alpha-\beta}{\alpha-\beta+2} L^{\alpha}
      +
      \mathcal{O} \left( \frac{L^{1+2\alpha -\beta}}{D} \right)
 \right)
       d\tau
       \notag\\
      &+
      \sqrt{D \, L^\beta} \ d W_\tau
     \ ,
\end{align}
 and the case $\alpha < \beta$:
 \begin{align}
\label{sup_tr_lv_tsa_sn_multN_alpha_sm_beta_simp}
  dL(\tau) =
  &\left(
    \gamma L^\alpha
      -  D (\alpha-\beta) L^{\beta-1} 
      +
      \mathcal{O} \left( \frac{D^2}{L^{2+\alpha-2\beta}} \right)
 \right)
       d\tau
\notag\\
      &+
      \sqrt{D \, L^\beta} \ d W_\tau
     \ .
\end{align}
Dependent on the powerlaw exponents $\alpha$ and $\beta$, the dynamics are either \textit{noise} or \textit{force} dominated. For $\alpha \ge \beta$,
the dynamics of Eq.~\eqref{sup_tr_lv_tsa_sn_multN_alpha_ge_beta_simp} are dominated by the noise, that is the $D$ dependent term $D L^{\beta-1}$.
The exact form of the force $f(L)$ and the assumed exponent $\alpha$ are thus irrelevant. The dynamics close to the target state behave as if only noisy fluctuations lead to absorption at the target. A detailed discussion of this case including analytical solutions is provided in \cite{lenner2023reverse2}.

For $\alpha < \beta-1$, and $\beta-1 \le \alpha < \beta$ in the joint limit of small noise and small $L$, the dynamics are \textit{force} dominated. The leading order term is given as the sign inverted forward force law. The next leading order term is proportional to the noise strength $D$. Its contribution to the dynamics is strictly positive.

\section{Statistics of homogeneous and heterogeneous TSA ensembles}
In this section, we formally examine how the contribution of parameter heterogeneity quantitatively impacts mean, variance and ensemble covariance of a heterogeneous TSA ensemble. While difficult to identify on the single trajectory level, heterogeneity can induce clear signatures in moments and correlation functions of the dynamics on the ensemble level.
We here denote the normalized ensemble distribution of the heterogeneous ensemble as $R^N(L,\tau;\eta)$. 
The $n$-th moment of the full heterogeneous ensemble 
\begin{align}
\label{eq:def_avg_over_cond_moments}
\langle (L(\tau) )^n \rangle_{R^N(L,\tau;\eta)}
&= 
  \int \int R^N(L,\tau;\eta) L^n dL \, d\eta
  \notag\\
  &=
  \int P(\eta) \left( \int R^N(L,\tau|\eta) L^n dL \right)  d\eta
\notag \\
  &=
\left[ \langle (L(\tau) )^n \rangle_{R^N(L,\tau|\eta)} \right]_{P(\eta)}
\end{align}
can be obtained by decomposing the full average into the homogeneous contribution subsequently averaged with respect to a variable parameter $\eta$. Here $\langle \cdot \rangle_{R^N(L,\tau|\eta)}$ denotes the homogeneous average with $\eta$ fixed and $\left[ \cdot \right]_{P(\eta)}$ the average over the heterogeneity parameter $\eta$. For readability we will generally drop the probability subscript of the respective ensemble average and use $\langle \cdot \rangle$ for coordinate averages and $\left[\cdot \right]$ for parametric averages.  We further denote the homogeneous mean and variance for a fixed parameter $\eta$ as 
\begin{align}
\overline{L}^{(\eta)}(\tau)&:=\langle L^{(\eta)}(\tau)  \rangle
\quad \mathrm{and}
\\
\sigma_L^{(\eta)2}(\tau)&:=
\langle (L^{(\eta)}(\tau))^2  \rangle
-
\langle L^{(\eta)}(\tau)  \rangle
\ .
\end{align}
Eq.~\eqref{eq:def_avg_over_cond_moments} is the formally correct expression to extend the analysis of homogeneous ensembles to the heterogeneous case. In its present form, it is however not informative about the systematic changes heterogeneity introduces to ensemble statistics. 

A more intuitive expression can be derived from a simple decomposition of random variables which belong to the same homogeneous ensemble denoted by $\eta$. We suggest to separate
\begin{align}
\label{trj_decomposition}
L^{(\eta)}(\tau) 
=
\overline{L}^{(\eta)}(\tau) + \sigma_L^{(\eta)}(\tau) \; l^{(\eta)}(\tau)
\end{align}
into its ensemble mean and a non-stationary stochastic process with variance $\sigma_L^{(\eta)2}(\tau)$, $\langle l^{(\eta)}(\tau) \rangle_{R^N(L,\tau|\eta)}=0$ and $\langle l_j^{(\eta) 2} (\tau)\rangle_{R^N(L,\tau|\eta)}=1$. Note that $l^{(\eta)}(\tau)$ does not have to be a Gaussian process.

 For the mean, the decomposition defined in Eq.~\eqref{trj_decomposition} leads to exactly the same result as the definition provided in Eq.~\eqref{eq:def_avg_over_cond_moments}. We find
\begin{align}
\underbrace{
\overline{L}(\tau)
}_\text{\textrm{ensemble mean}}
=
\underbrace{
\left[ \overline{L}^{(\eta)}(\tau) \right]
}_\text{\textrm{mean of means}}
 \ .
\end{align}
In some instances, including the ones discussed below, the mean of the heterogeneous ensemble will only be a re-scaled version of the homogeneous case. The ensemble mean is thus of limited value for distinguishing between heterogeneous and homogeneous dynamics.

The variance is similarly obtained as the mean. Using the definition in Eq.~\eqref{trj_decomposition} and after some rearrangements we find
\begin{align}
\underbrace{
\sigma^2_L(\tau)
}_\text{\textrm{ensemble variance}}
&=
\left[ ( \sigma^{(\eta)}_L )^2(\tau) \right]
+
\left[ (\overline{L}^{(\eta)})^2(\tau) \right]
-
\left[ \overline{L}^{(\eta)}(\tau) \right]^2
\notag \\
&=
\underbrace{
\left[ (\sigma^{(\eta)}_L)^2(\tau) \right]
}_\text{\textrm{mean variance}}
+
\underbrace{
\left[ (\delta \overline{L}^{(\eta)})^2(\tau) \right]
}_\text{\textrm{variance of means}}
\ .
\end{align}
This decomposition is typically called the law of total variance. The variance of heterogeneous ensembles can thus be decomposed into the mean variance of uniform subsample and the variance of the subsample means. 

The first term is simply the variance of the homogeneous case averaged with respect to its parameter dependence. If the parameter dependence occurs only as a scaling factor, the functional form of this contribution to the variance will be unchanged and only the scaling in the heterogeneous case will be different.

The second contribution to the variance is exclusively due to the heterogeneity in the ensemble as each subsample contributes due to its respective mean. This term vanishes for homogeneous ensembles. In many instances, this term can change the functional form of the variance. 

The two-times-covariance can be decomposed in a similar fashion. We find that the ensemble covariance
\begin{align}
\underbrace{
C_L(\tau,\tau')
}_\text{\textrm{ensemble covariance}}
&=
\underbrace{
\left[ C^{(\eta)}_L(\tau,\tau') \right]
}_\text{\textrm{mean covariance}}
\notag \\
&+
\underbrace{
\left[ \overline{L}^{(\eta)}(\tau) \overline{L}^{(\eta)}(\tau') \right]
-
\overline{L}(\tau) \overline{L}(\tau')
}_\text{\textrm{covariance of means}}
\end{align}
decomposes into the mean covariance of the homogeneous subsamples and the covariance of the subsample means. This decomposition is typically known as the law of total covariance. The first contributing term may often only be a "re-normalized" version of the covariance of homogeneous dynamics. For these cases only the value of the effective parameter changes. 

The contribution representing the covariance of the subsample means can quantitatively change the ensemble covariance, by introducing non–decaying long-term correlations, that is "quenched fluctuations". We will see below, that such quenched fluctuations are a telltale signature of ensemble heterogeneity.

In the following sections we study the noise dominated and force dominated case. For both we consider the change in the ensemble statistics due to sample to sample variability in the diffusion constant $D$ and in the strength of the force term $\gamma$.

\subsection{The homogeneous case}
We first discuss the case of homogeneous dynamics, recapitulating results from \cite{lenner2023reverse1,lenner2023reverse2}. This serves to clarify the contrast between noise dominated and force dominated target state approach, and to slightly rewrite results to make the extension to heterogeneity apparent. In the following section, we will then discuss heterogeneity and explore its influence on the same statistics that are presented here.

\subsubsection{The noise dominated case}
\begin{figure}[ht]
 \centerline{\includegraphics[width=\linewidth]
 {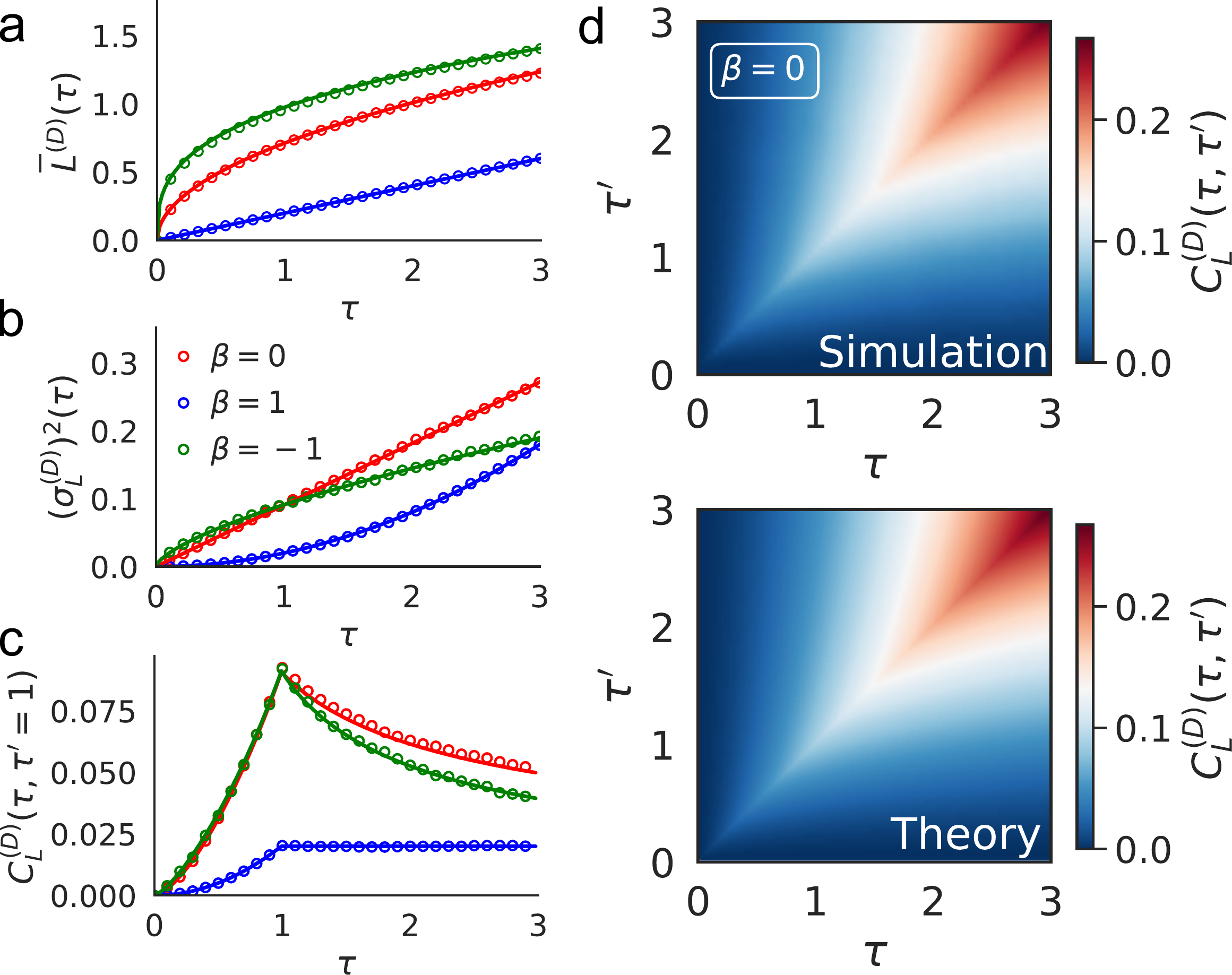}}
 \caption{
 \textbf{Noise dominated TSA dynamics of homogeneous samples are distinguishable for different power law noise models.}
 Shown is a comparison of mean \textbf{(a)}, variance \textbf{(b)} and "covariance-cuts" \textbf{(c)} for dynamics with identical force law $f(L)=-\gamma L^\alpha$ (here $\alpha=-1$),
 but different multiplicative noise $D(L)=D L^\beta$ with $\beta=0,1,-1$. We observe perfect agreement between our theory (lines) and simulations (circles).
The comparison of theory and simulation for the full covariance is shown for $\beta=0$ in \textbf{(d)}. The other cases can be found in the supplementary information.
The ensemble statistics of the forward dynamics has been simulated with 20000 trajectories starting at $\widehat{L}_0=3$. The analytic expressions for the mean, variance and covariance are stated in Eq.~\eqref{MultNoiseDriven_mean} and Eq.~\eqref{MultNoiseDriven_var} and Eq.~\eqref{MultNoiseDriven_cov}. The parameters are $\gamma=1$ and $D=0.2$.
 }\label{fig:noisedominated_meanvarcov_cuts}
 \end{figure}

The noise dominated regime occurs for the parameter regime $\alpha \ge \beta$.
In our previous work \cite{lenner2023reverse2} we found for the mean of in $D$ homogeneous TSA ensembles with multiplicative noise
\begin{align}
\label{MultNoiseDriven_mean}
\overline{L}^{(D)}(\tau)
&=
 \frac{2^{\frac{\beta -1}{\beta -2}} (2-\beta )^{-\frac{2}{\beta -2}} \Gamma \left(-\frac{2}{\beta -2}\right) }{\Gamma \left(\frac{1}{2-\beta }\right)}
 (D\tau)^{\frac{1}{2-\beta }}
 \\
 &=: D^{\frac{1}{2-\beta }} M_1(\tau)
\end{align}
and for the variance 
\begin{align}
\label{MultNoiseDriven_var}
(\sigma_L^{(D)})^2(\tau)
=
&(2-\beta )^{-\frac{4}{\beta -2}} \Bigg(3\cdot 4^{\frac{1}{\beta -2}}
   \frac{\Gamma \left(-\frac{3}{\beta -2}\right)}{\Gamma \left(\frac{1}{2-\beta }\right)}
   \notag\\
   &-
   4^{\frac{\beta -1}{\beta -2}} 
   \frac{\Gamma \left(-\frac{2}{\beta-2}\right)^2}{\Gamma \left(\frac{1}{2-\beta }\right)^2}
   \Bigg)
   (D \tau)^{\frac{1}{1-\beta/2}}
   \\
   =:&
   D^{\frac{2}{2-\beta}} \left(M_2(\tau) - M_1^2(\tau) \right)
   \\
    =:& D^{\frac{2}{2-\beta}} S^2(\tau)
    \ .
\end{align}
To make the scaling with $D$ apparent, we introduced the scaling factor free moments $M_i(\tau)$ and variance $S^2(\tau)$. Mean and variance evaluated for  $\beta=-1,0,1$ are shown in Fig.~\ref{fig:noisedominated_meanvarcov_cuts}. They are selected to show the change in the variance from a concave dependence on $\tau$ for $\beta<0$ to a convex form for $\beta>0$. Interesting for system identification, the coefficient of variation, defined as ratio of the noise level to the mean $CV(\tau)=\frac{\sigma_L(\tau)}{\overline{L}(\tau)}$, is constant for all $\beta$. Only homogeneous force dominated dynamics with $\beta=\alpha+1$ can also show such a behavior.

We did not find a closed form solution of the covariance for general $\beta$. For a given $\beta$, however closed form expressions can be obtained from the joint probability distribution of the noise dominated case which we state in the supplementary information. As an example case we state the intriguing case with $\beta=1$. We find
\begin{align}
\label{MultNoiseDriven_cov}
C_L^{(D)}(\tau,\tau') &= \frac{D^2}{2} \min(\tau,\tau')^2 
\\
&= (\sigma_L^{(D)})^2(\min(\tau,\tau'))
\qquad (\beta=1)
\ .
\end{align}
For $\beta=1$ the covariance exclusively depends on the shorter time to completion say $\tau'$, and does not decay for larger times $\tau$. In forward time, scale free processes, such as random walk dynamics with boundaries at $\pm \infty$ and constant $D$, that is $\beta$=0, show such a behavior. In Fig.~\ref{fig:noisedominated_meanvarcov_cuts}, we show the covariance case with $\beta=0$. We additionally show the "decay" of the covariance with $\tau$ for a fixed $\tau'=\tau_\mathrm{fix}$ and $\beta \in \{-1,0,1\}$. With increasing $\beta$ the decay length of the covariance becomes visibly longer until it reverts to the increasing case for $\beta>1$. 

To demonstrate the reliability of our approach, we additionally compare our theoretical results of mean, variance and covariance to simulations of the dynamics in forward time, subsequently target state aligned, and evaluated with respect to their ensemble statistics. We find excellent agreement, as shown in Fig.~\ref{fig:noisedominated_meanvarcov_cuts}

\subsubsection{The force dominated case}
\begin{figure}[ht]
\centerline{\includegraphics[width=\linewidth]
{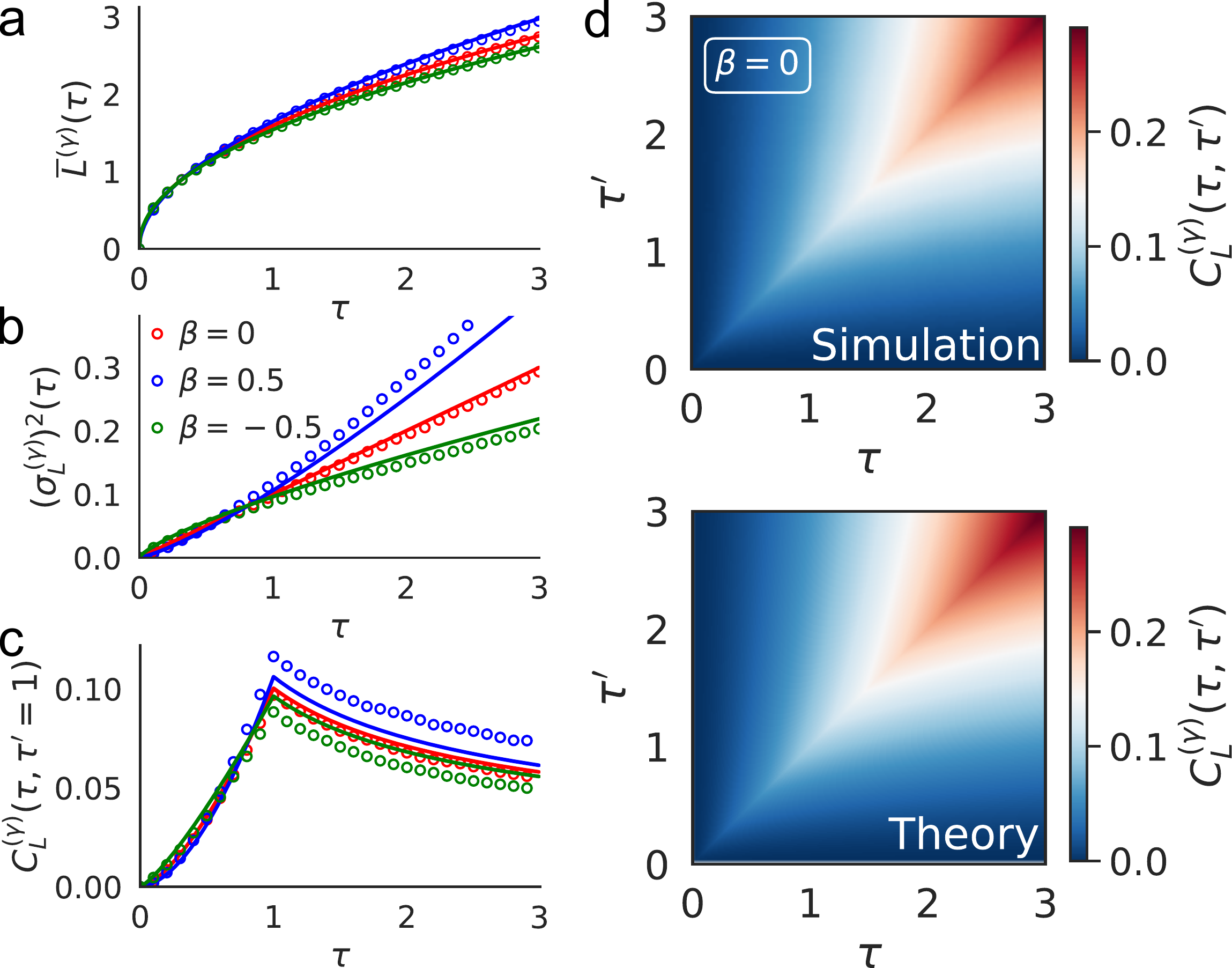}}
\caption{
\textbf{Force dominated TSA dynamics of homogeneous samples are distinguishable with respect to different power law forces and noise models.} We compare dynamics with identical force laws $f(L)=-\gamma L^\alpha$ (here $\alpha=-1$),
 but different multiplicative noise $D(L)=D L^\beta$ with $\beta=0,0.5,-0.5$.
 We find almost perfect agreement for mean \textbf{(a)}, variance \textbf{(b)} and "covariance-cuts" \textbf{(c)} of simulations (circles) and our theory (lines) for the case with $\alpha =\beta-1$. For this case, the order $D$ contribution to the force is of the same functional form as the force term (Eq.~\eqref{sup_tr_lv_tsa_sn_multN_alpha_sm_beta_simp}). The observed deviations increase in the variance (and the variance dependent part of the covariance) the further we deviate from this equality. We show the comparison of the full covariance for the case $\beta=0$ in \textbf{(d)}. The other cases can be found in the supplementary information.
 The ensemble statistics of the forward dynamics has been simulated with 20000  trajectories that start at $\widehat{L}_0=20$. The analytic expressions for the mean, variance and covariance are stated in Eq.~\eqref{sup_mean_sn_mult} and Eq.~\eqref{MultForceDriven_var} and Eq.~\eqref{sup_sn_cov}. Parameters are $\gamma=1$ and $D=0.2$.
 }\label{fig:ForceDriveTSA_meanVarCov_cuts}
 \end{figure}

Force dominated target state arrival occurs for $\alpha<\beta-1$ and for $-1<\alpha-\beta<0$ under the additional constraint of small $D$ -
to be precise for $D\to 0$ as $L \to 0$.
We only consider cases with $\alpha<1$, where the deterministic dynamics terminate in finite time. 
In \cite{lenner2023reverse2}, we derive small noise moments from Eq.~\eqref{sup_tr_lv_tsa_sn_multN_alpha_sm_beta_simp} up to order $D$.
In the multiplicative noise case, we found for the mean 
\begin{align}
\label{sup_mean_sn_mult}
 &\overline{L}^{(\gamma)}(\tau)
 =
   ((1-\alpha ) \gamma  \tau)^{\frac{1}{1-\alpha }}
   \notag\\
&\quad \ +
\frac{D}{\gamma}
   \frac{ \left(7 \alpha ^2-\alpha  (8 \beta +3)+2 \beta  (\beta +1)\right) ((1-\alpha) \gamma \tau)^{\frac{\beta - \alpha}{1 - \alpha}}}{2  (\beta - 2 \alpha) (1 - 3 \alpha +\beta)}
   \notag \\
  &\qquad \quad
  =: \gamma^{\frac{1}{1-\alpha}} \tilde{L}_0(\tau) 
  + D \; \gamma^{\frac{\beta -1}{1- \alpha}} \langle \tilde{L}_2(\tau) \rangle
  \ .
\end{align}
For use in the later discussion on heterogeneity, we explicitly separated the parametric $\gamma$-dependence from the two expansion terms of order zero and two. We denote these residual terms as $\tilde{L}_i$, where $i$ denotes the expansion order. The term with $\tilde{L}_0$ is the deterministic solution of the process. Terms of order one do not contribute as we detail in the supplementary information. The lowest order $\beta$ dependent term is of second order. This suggests that the functional form of the mean is largely dominated by $\alpha$. We show this exemplary for the case $\alpha=-1$ in Fig,~\ref{fig:ForceDriveTSA_meanVarCov_cuts}.

The variance up to order $D$ is equivalent to the average over the squared first order term of the small noise expansion. We find
\begin{align}
\label{MultForceDriven_var}
(\sigma_L^{(\gamma)})^2(\tau)
 &=  
   \frac{D}{\gamma}
   \frac{ ((1 -\alpha)\gamma \tau)^{\frac{1-\alpha+\beta }{1-\alpha }}}{1-3 \alpha
   +\beta}
   \\
      &=:   
D \; \gamma^\frac{\beta }{1-\alpha } \langle \tilde{L}_1^2(\tau) \rangle
\ .
\end{align}

Keeping in mind that $\alpha<1$ holds, the slope of the variance
is determined by $\beta$. For $\beta >0$ the variance becomes convex, for $\beta <0$ it is concave and for $\beta=0$ it is linear. Three cases depicting this behavior are shown in Fig.~\ref{fig:ForceDriveTSA_meanVarCov_cuts}.

The two-time covariance with multiplicative noise can be expressed as power law scaled versions of the variance. We find
\begin{align}
 \label{sup_sn_cov}
 C_L^{(\gamma)}(\tau,\tau')
   &=
   D \; \gamma^\frac{\beta }{1-\alpha }  
\left(
\frac{\min(\tau,\tau')}{\max(\tau,\tau')}
\right)^{\frac{\alpha }{\alpha -1}}
\; \langle \tilde{L}_1^2(\min(\tau,\tau') ) \rangle
   \notag
\\
&=: D \; \gamma^\frac{\beta }{1-\alpha } c(\tau,\tau')
\ ,
\end{align}
where the last line introduces the $\gamma$ and $D$ independent version of the covariance $c(\tau,\tau')$.
For fixed $\tau'$ with $\tau>\tau'$, the change of the covariance with $\tau$ depends exclusively on $\alpha$ and is independent of $\beta$. For this choice of $\tau$ and $\tau'$, the covariance always decays with large $\tau$ for  $\alpha<0$ and increases for $\alpha>0$. Note that the latter case can strictly only occur if $\beta>1$ and for very small $D$ if $\beta > 0$.
To demonstrate the validity of our calculations, we show the covariance for $\beta=0$ and $\alpha=-1$ in Fig.~\ref{fig:ForceDriveTSA_meanVarCov_cuts}. We also show "covariance-cuts" with one time axis $\tau'=\tau_\mathrm{fix}$ kept fix. The effect of $\beta$ is visible yet the dominant shaping parameter of the covariance is clearly $\alpha$.

For all the shown theoretical curves, we also provide results from simulations of the underlying process in forward time, which we analyzed with respect to their TSA ensemble statistics. We find good agreement between moments obtained from simulations and moments from our small noise approximation.


\subsection{The heterogeneous case}
In this section, we will investigate TSA dynamics including parameter heterogeneity.

\subsubsection{The noise dominated case}
In the \textit{noise} dominated case and to first order, any form of heterogeneity in the force law will be undetectable. The ensemble statistics exclusively depend on the diffusion term, that is on the power law exponent $\beta$. Heterogeneity in noise dominated system is therefore exclusively to be found as heterogeneity in the diffusion constant $D$. 

For these moments, the heterogeneous generalization is easily obtained using Eq.~\eqref{fig:ForceDriveTSA_meanVarCov_cuts}. We find for the mean and variance of the heterogeneous ensemble
\begin{align}
\label{eq:noiseDom_het_mean}
\overline{L}(\tau)
&=
 \left[ D^{\frac{1}{2-\beta }} \right] M_1(\tau) 
 \quad \mathrm{and}
 \\
 (\sigma_L)^2(\tau)
 &= 
 \left[ D^{\frac{2}{2-\beta }} \right] M_2(\tau)
 -
 \left[ D^{\frac{1}{2-\beta }} \right]^2 M_1^2(\tau)
 \ .
\end{align}
Adding and subtracting the term 
$\left[ D^{\frac{2}{2-\beta }} \right] M_1^2(\tau)$ to the variance we find
\begin{align}
 \label{eq:noiseDom_het_var}
 (\sigma_L)^2(\tau)
 &= 
 \left[ D^{\frac{2}{2-\beta }} \right] S^2(\tau)
 + \mathrm{Var} \left[D^{\frac{1}{2-\beta }} \right] M_1^2(\tau)
\end{align}
which is equivalent to the mean over the homogeneous variance plus the variance of the means.

\begin{figure}[ht]
 \centerline{\includegraphics[width=\linewidth]
 {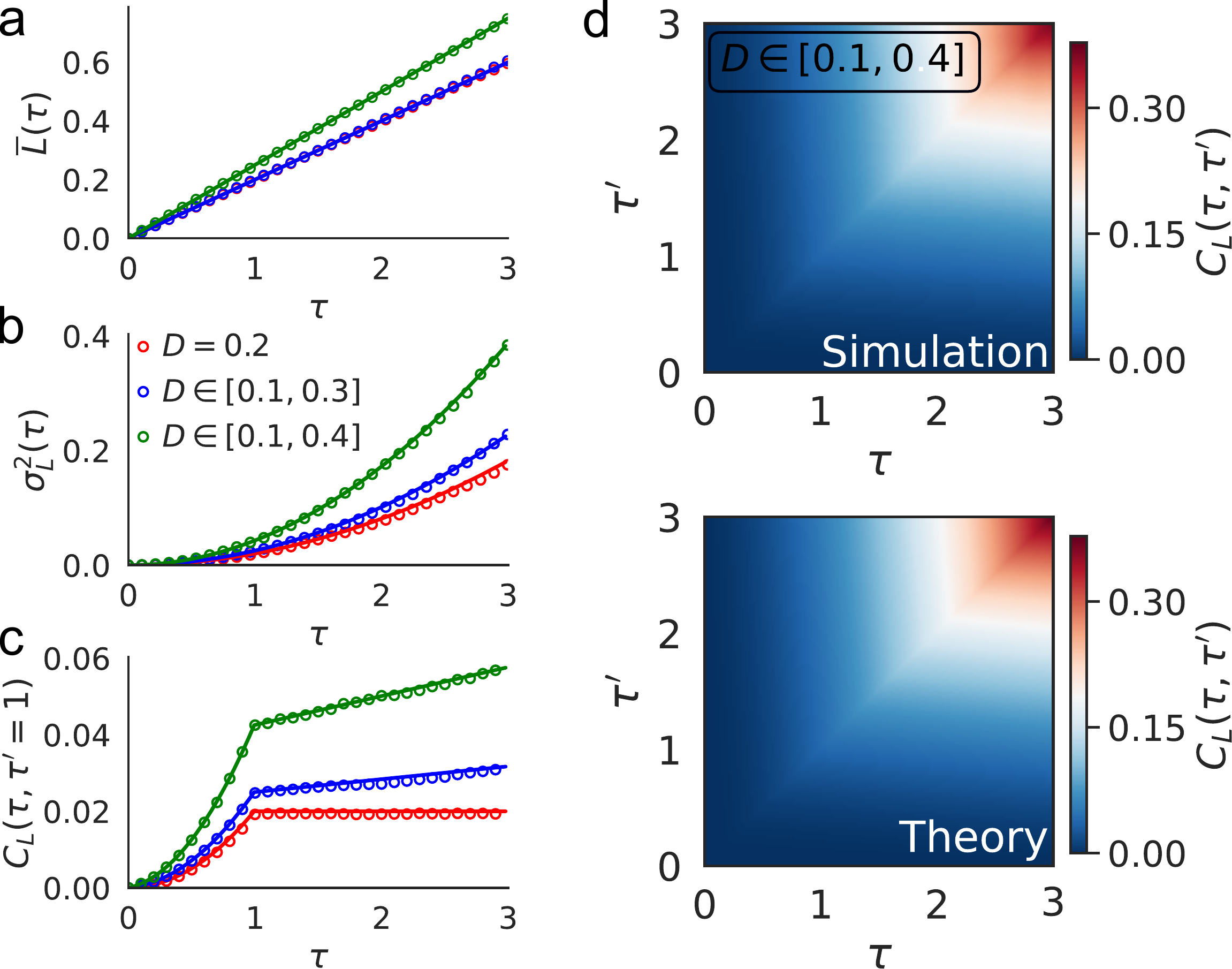}}
 \caption{
 \textbf{Heterogeneity in the diffusion constant $D$ of noise dominated TSA dynamics is detectable in the covariance.}
 We compare the homogeneous case with $D(L)= D L$  studied in Fig.~\ref{fig:noisedominated_meanvarcov_cuts} ($D=0.2$), to the heterogeneous version with $D$ randomly drawn from a fixed interval for each sample path realization. 
 We chose $D\in \left[0.1 , 0.3 \right]$ and $D\in \left[0.1 , 0.4 \right]$.
 While mean \textbf{(a)} and variance \textbf{(b)} show no signature of heterogeneity, is the covariance \textbf{(c)}, \textbf{(d)} a clear readout of heterogeneity. Starting from constant non-decaying covariance cuts for $\tau>\tau_\mathrm{fix}$ in the homogeneous case, these lines gain increasing slopes with higher heterogeneity. The other covariance cases are shown in the supplementary information.
Simulations (circles) and theory (lines) are, as already in the homogeneous case, in excellent agreement.
The statistics of the forward dynamics are based on 15000  trajectories that start at $\widehat{L}_0=6$. The analytic expressions for the mean, variance and covariance are stated in Eq.~\eqref{eq:noiseDom_het_mean}, Eq.~\eqref{eq:noiseDom_het_var}, Eq.~\eqref{eq:noiseDom_het_cov} and Eq.~\eqref{eq:noiseDom_het_cov_betaOne}. The parameter $\gamma=0$ was chosen.
 }\label{fig:noisedominated_het}
 \end{figure}

Interestingly, both mean and variance of noise dominated dynamics are insensitive to heterogeneity with respect to their functional dependency on $\tau$. We find $\overline{L}(\tau)\sim \tau^{\frac{1}{2-\beta}}$ and $(\sigma_L)^2(\tau) \sim \tau^{\frac{2}{2-\beta}}$. 
We show this behavior in Fig.~\ref{fig:noisedominated_het}.
This implies that the $CV$ for heterogeneous dynamics is constant with time analogously to the homogeneous case. 

From our discussion above we know, that the covariance decomposes in the mean of the homogeneous case and the covariance of means. Using that the homogeneous covariance scales with the same factor as the homogeneous variance, defined in Eq.~\eqref{MultNoiseDriven_var}, we write
\begin{align}
 \label{eq:noiseDom_het_cov}
C_L(\tau,\tau')
=
    \left[D^{\frac{2}{2-\beta}}\right] c(\tau,\tau')
    + 
\mathrm{Var}\left[D^{\frac{1}{2-\beta}} \right] M_1(\tau) M_1(\tau')
\ .
\end{align}
For the case $\beta=1$, which we studied above for the homogenous case, we find
\begin{align}
 \label{eq:noiseDom_het_cov_betaOne}
C_L(\tau,\tau')
=
    \left[D^2\right] \frac{\min(\tau,\tau')^2}{2}
    + 
\mathrm{Var}\left[D \right] \tau \tau'
\ .
\end{align}
For this specific case, heterogenity changes the with $\tau$ non-decaying homogenous covariance into a case where the covariance increases with $\tau$. We show this effect in Fig.~\ref{fig:noisedominated_het}. In general, adding heterogeneity leads to lesser and lesser decaying covariance. For larger $\beta$-values the introduction of heterogeneity can even lead to with $\tau$ increasing covariance (see Fig.~\ref{fig:noisedominated_het}). 
Heterogeneity in noise dominated dynamics is thus mostly detectable with the help of the two-time covariance.

\subsubsection{The force dominated case}
The statistics of force dominated dynamics are sensitive to parametric heterogeneity both in the noise and the force law.
The generalization to heterogeneous moments is slightly more involved than the noise dominated case. We again apply Eq.~\eqref{fig:ForceDriveTSA_meanVarCov_cuts} to generalize our results for homogeneous dynamics to the heterogeneous case. As the force dominated case is however built on an expansion, and not on exact moments as for the noise dominated case, the introduction of heterogeneity will include expansion terms of different order. A detailed account is given in the supplementary information.

We first study the case of sample to sample heterogeneity in the diffusion constant. We find for the mean
\begin{align}
\label{sup_mean_sn_mult_Dhet}
 &\overline{L}(\tau)
  =: 
  \gamma^{\frac{1}{1-\alpha}} \tilde{L}_0(\tau) 
  + 
  \left[D\right] \; \gamma^{\frac{\beta -1}{1- \alpha}} \langle \tilde{L}_2(\tau) \rangle
\ ,
\end{align}
variance 
\begin{align}
\label{MultForceDriven_var_Dhet}
(\sigma_L^2(\tau)
      &=:   
\left[D\right] \; \gamma^\frac{\beta }{1-\alpha } \langle \tilde{L}_1^2(\tau) \rangle
+ \ \cdots
\notag \\
&\quad \ + 
\mathrm{Var}\left[ D \right] \gamma^\frac{2\beta-2}{1-\alpha}
\langle \tilde{L}_2(\tau) \rangle^2
\end{align}
and two-time covariance
\begin{align}
 \label{sup_sn_cov_Dhet}
 C_L(\tau,\tau')
&=: \left[D \right] \; \gamma^\frac{\beta }{1-\alpha } c(\tau,\tau')
+ \ \cdots
\notag \\
&\quad \ +
\mathrm{Var}\left[ D \right] \gamma^\frac{2\beta-2}{1-\alpha}
  \langle \tilde{L}_2(\tau) \rangle
  \langle \tilde{L}_2(\tau') \rangle
  \ . 
\end{align}
Up to order $D$, $D$-Heterogeneity appears as average over the set of diffusion constants which are present in the full ensemble. The  "$\cdots$"  represent all other terms of  order $D^{3/2}$ or higher that were not considered. The expected contribution due the variance of the means, readily obtainable from Eq.~\eqref{sup_mean_sn_mult}, shows up if the expansion is extended to include order $D^2$ terms (which implies to take the expansion to forth order). 

\begin{figure*}[ht]
 \centerline{\includegraphics[width=\linewidth]
 {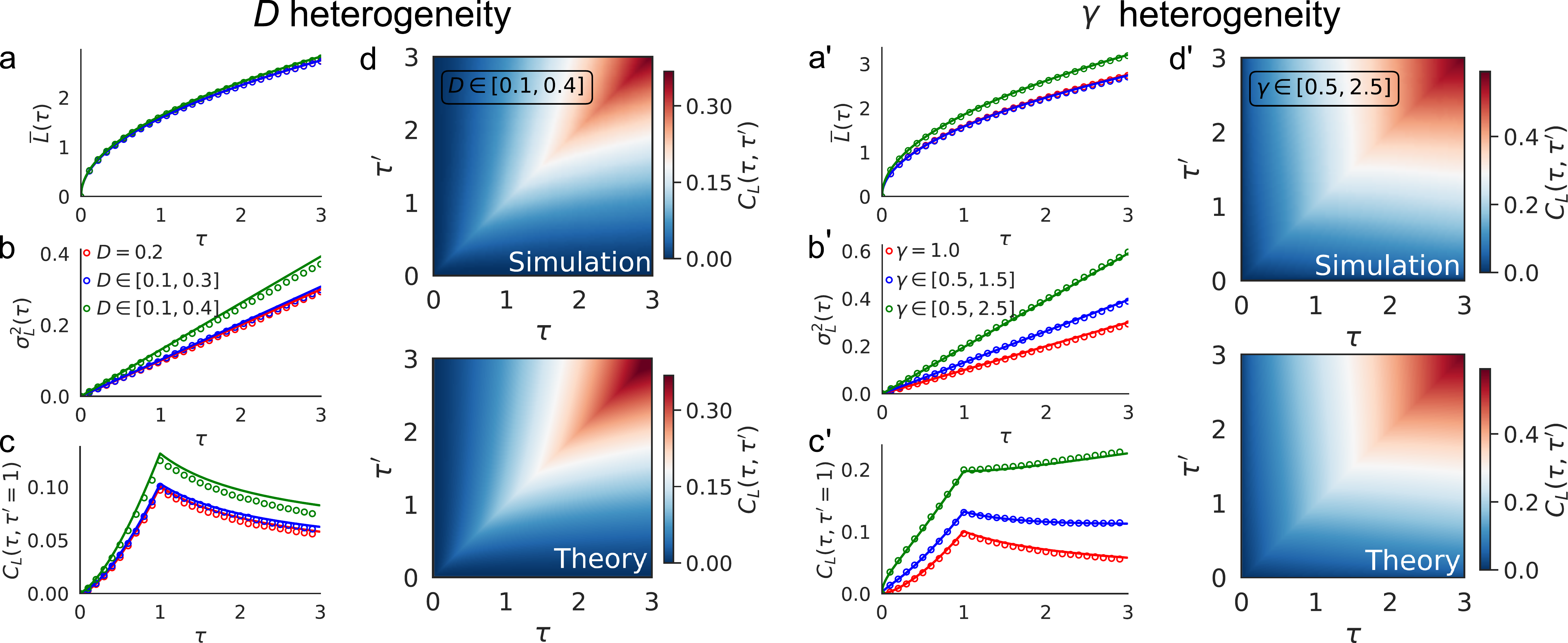}}
 \caption{
 \textbf{Parametric heterogeneity of force dominated TSA dynamics is revealed in the covariance.} We compare the ensemble statistics of homogeneous dynamics  with $f(L)=-\frac{\gamma}{L}$ and $D(L)= D$ to the heterogeneous case with $D$ (\textbf{(a)},\textbf{(b)},\textbf{(c)},\textbf{(d)})  and $\gamma$ (\textbf{(a')},\textbf{(b')},\textbf{(c')},\textbf{(d')}) randomly drawn from a fixed interval for each sample path realization. For $D$-heterogeneity the chosen intervals are $D \in [0.1,0.3]$ and $D \in [0.1,0.4]$. For $\gamma$-heterogeneity we chose $\gamma \in [0.5,1.5]$ and $\gamma \in [0.5,2.5]$. The homogeneous case, which serves as a reference for both cases, uses $D=0.2$ and $\gamma=1.0$. In general, we find simulations (circles) and our small noise theory for heterogeneous dynamics (lines) are in good agreement.
 In the here studied small noise regime, $D$-heterogeneity (\textbf{(a)},\textbf{(b)},\textbf{(c)},\textbf{(d)}) is naturally less prominent. We find small difference between homogeneous and heterogeneous dynamics both in the variance (\textbf{(b)}) and in the covariance (\textbf{(c)},\textbf{(d)}). For larger noise strength the contribution of the variance of means becomes more relevant.
 For cases with $\gamma$-heterogeneity (\textbf{(a')},\textbf{(b')},\textbf{(c')},\textbf{(d')}), we find that heterogeneity induces changes in mean, variance and covariance.
The mean \textbf{(a')} changes depending on the mean of the chosen interval. The variance \textbf{(b')} increases with increasing heterogeneity. However, only the covariance \textbf{(c')},\textbf{(d')} changes its functional form with increasing heterogeneity. The covariance is thus a clear marker for the degree of $\gamma$-heterogeneity of the studied dynamics. The respective covariance cases, which here are only represented as "covariance-cuts", are shown as 2d-plots in the supplementary information.
 The statistics of the target state aligned forward simulations are based on 20000 trajectories that start at $\widehat{L}_0=20$. The analytic expression for the mean, variance and covariance for the $D$-heterogeneous case are stated in Eq.~\eqref{sup_mean_sn_mult_Dhet}, Eq.~\eqref{MultForceDriven_var_Dhet} and Eq.~\eqref{sup_sn_cov_Dhet}. The respective equations for the $\gamma$-heterogeneous case are stated in Eq.~\eqref{mean_het_gamma}, Eq.~\eqref{var_het_gamma} and Eq.~\eqref{cov_het_gamma}.
 %
 %
 %
}\label{fig:ForceDrivenTSA_het_mean_var_cov}
\end{figure*}

In Fig. \ref{fig:ForceDrivenTSA_het_mean_var_cov}, we show a comparison between homogeneous dynamics and dynamics with sample to sample variability in $D$. In the small noise regime and for dynamics where the averaged $D$ equals the homogeneous $D$ we find small differences due to the variance of the means.
In the limit of small noise, $D$ heterogeneity therefore occurs approximately as an effective, that is averaged, diffusion constant. For larger noise strength the contribution due to the variance of the means will become more relevant.

In the remainder of this section we consider the case of sample to sample variability in the force strength $\gamma$. The aforementioned coupling of higher order terms due to the introduction of parametric heterogeneity becomes relevant for variance and covariance.
For the mean we find
\begin{align}
\label{mean_het_gamma}
\overline{L}(\tau)
&=
  \left[\gamma^{\frac{1}{1-\alpha}}\right] \tilde{L}_0(\tau) 
  + D \left[\gamma^{\frac{\beta -1}{1- \alpha}}\right] \langle \tilde{L}_2(\tau) \rangle
  \ ,
\end{align}
which is the very same expression as for the homogeneous case (Eq.~\eqref{sup_mean_sn_mult}), however with averages over the $\gamma$-dependent terms.

For the variance we find 
\begin{align}
\label{var_het_gamma}
\sigma_L^2(\tau)
&=
 D 
\left[
\gamma^\frac{\beta }{1-\alpha }
\right]
\langle \tilde{L}_1^2(\tau) \rangle \
\times \notag \\
&\times
\left(
1+ 
    \left(
        1- \frac{\left[ \gamma ^{\frac{1}{1-\alpha }} \right] 
        }{\left[ \gamma ^{\frac{\beta }{1-\alpha }} \right]}
        \left[ 
        \gamma ^{\frac{\beta-1}{1-\alpha}}
        \right]
    \right) h(\alpha,\beta)
\right)
\notag \\
&+
\mathrm{Var}\left[\gamma^{\frac{1}{1-\alpha}} \right] 
\tilde{L}_0^2(\tau)
\end{align}
with
\begin{align}
h(\alpha,\beta)
:=
    \frac{ 7 \alpha ^2-\alpha  (8 \beta +3)+2 \beta  (\beta +1)}{\beta - 2 \alpha }
    \ .
\end{align}
The order $D$ behavior of the homogeneous multiplicative case is preserved, that is, a convex $\tau$-dependence for $\beta >0$, and a concave curve for $\beta<0$. All modifications due to heterogeneity in $\gamma$ are only in the prefactor and do not change the dependence of this term on $\tau$. 
Note that the  multiplicative factor stated in the second line of Eq.~\eqref{var_het_gamma} evaluates to one for $\beta=1$.
Comparing terms, the first term contributing to the heterogeneous variance is thus the mean of the homogeneous variance times a correction due to the coupling of higher and lower order terms in the expansion, which only contribute in the heterogeneous case (see supplementary information).

The second contribution to the variance is of zero-th order in $D$ and only contributes for non-negligible heterogeneity. It is comprised of the variance of $\gamma^{\frac{1}{1-\alpha}}$ times the squared solution of the deterministic dynamics. For $\alpha < -1$ it adds a concave in $\tau$ contribution to the variance, for $\alpha >-1$ the contribution is convex and linear in between. In the terminology from above, this term is the contribution due to the variance of the means.
We show three cases of increasing heterogeneity in Fig.~\ref{fig:ForceDrivenTSA_het_mean_var_cov}. In the specific case shown, the behavior of the mean can stay unchanged if the interval is evenly increased to both larger and smaller $\gamma$ values. In this case, the variance increases continuously with an increase in the interval.

The two-time covariance
\begin{align}
\label{cov_het_gamma}
 &C_L(\tau,\tau') = 
D \; \left[ \gamma^\frac{\beta }{1-\alpha } \right] \; 
c(\tau,\tau') 
\notag \\
&+
D \; \left[ \gamma^\frac{\beta }{1-\alpha } \right] \;
    \left(
        1- \frac{\left[ \gamma ^{\frac{1}{1-\alpha }} \right] 
        }{\left[ \gamma ^{\frac{\beta }{1-\alpha }} \right]}
        \left[ 
        \gamma ^{\frac{\beta-1}{1-\alpha}}
        \right]
    \right)
h(\alpha,\beta)
\times
\notag \\
& \qquad \times
\frac{1}{2}
\left(
\langle \tilde{L}_1^2(\tau) \rangle
    \left(\frac{\tau'}{\tau} \right)^\frac{\beta-\alpha}{1-\alpha}
    +
\langle \tilde{L}_1^2(\tau') \rangle
    \left(\frac{\tau}{\tau'} \right)^\frac{\beta-\alpha}{1-\alpha}    
\right)
\notag \\
 &+
 \mathrm{Var}\left[\gamma^{\frac{1}{1-\alpha}} \right]
 \tilde{L}_0(\tau) \; \tilde{L}_0(\tau')
\end{align}
is most telling about the heterogeneity of the dynamics. At its core it is comprised of two parts, which are the covariance due to random fluctuations (first line) and the covariance contribution due to the heterogeneity of the deterministic dynamics (last line). The cross-term between both terms (second and third line) is mostly negligible and zero for $\beta=1$. 

Fundamentally, we again find the partitioning into the mean of the covariance, and the covariance of means. In particular the latter leads to the interesting phenomenon of a with $\tau$ (for $\tau > \tau'$) non decaying covariance for sufficient heterogeneity. We show this behavior exemplary in Fig.~\ref{fig:ForceDrivenTSA_het_mean_var_cov}.


\section{Discussion}
The TSA approach allows to analyse the dynamics of a system independent of initial conditions and knowledge of preceding dynamics. In this paper we generalize the TSA framework recently introduced in Lenner et.~al.~\cite{lenner2023reverse1,lenner2023reverse2} for constant and multiplicative noise, to heterogeneous dynamics. While target state alignment provides a way to analyze otherwise hardly accessible ensembles, it also introduces pseudo forces due to alignment. Our framework provides the means not only to separate these pseudo-forces from the dynamics, but also to characterize the forward process based on an ensemble we study in reverse time.

For general TSA dynamics, we distinguish between noise-dominated and force-dominated target state arrival, for which we provide analytical expressions for mean, variance and two-time covariance. From our previous work on TSA dynamics, we know that straightforward inference based on the mean can lead to incorrect assignment of the actual underlying force laws due to the presence of pseudo forces. Because of this complication, the mean is only a good proxy for force dominated dynamics. It provides no information about the force law in the noise dominated case. 

For the practical inference of the underlying force law from data it is of great importance to disentangle the effects of intrinsic state-dependent noise and heterogeneity in the effective parameters. Using our analytical expressions for the variance, the form of the state dependent noise can be directly read off for both homogeneous and heterogeneous dynamics. This holds for both the force and the noise dominated cases. Therefore, the intrinsic state-dependent noise can be confidently identified within our approach.

Using the law of total variance and total covariance, we show how our results for the homogeneous ensemble statistics can be generalized to the heterogeneous case. Both the results for the homogeneous variance and covariance reappear in the heterogeneous case as an averaged version with respect to their parameter heterogeneity. In addition, we find a term proportional to the variance of the means of the parametrically distinguishable subsamples. This latter term can alter the functional form of both the variance and the covariance and thus be used to identify the degree of heterogeneity in the system under study. We confirm our theoretical results with numerical simulations performed in forward time and subsequently analyzed as a TSA ensemble.

To classify experimentally observed TSA dynamics, the following scheme can be applied: i) identify whether the dynamics are noise-dominated using the coefficient of variation. If not, ii) assume that the dynamics are force dominated and infer the power law exponent $\alpha$ of the force from the mean and the noise exponent $\beta$ from the variance. iii) Finally, the covariance reveals whether a process is homogeneous or heterogeneous.

In summary, our framework provides the means to dissect TSA ensembles with respect to their dynamics, noise statistics and heterogeneity. It can be used both for classification of target state arrival and parameter inference.


\begin{acknowledgments}
We thank Erik Schultheis and the Wolf group for stimulating discussions and proofreading of the manuscript. This work was supported by the German Research Foundation (Deutsche Forschungsgemeinschaft, DFG) through FOR 1756, SPP 1782, SFB 1528, SFB 889, SFB 1286, SPP 2205, DFG 436260547 in relation to NeuroNex (National Science Foundation 2015276) \& under Germany’s Excellence Strategy - EXC 2067/1- 390729940; by the Leibniz Association (project K265/2019); and by the Niedersächsisches Vorab of the VolkswagenStiftung through the Göttingen Campus Institute for Dynamics of Biological Networks.
\end{acknowledgments}


\bibliography{literature}

\providecommand{\noopsort}[1]{}\providecommand{\singleletter}[1]{#1}%
\begin{thebibliography}{32}%
\makeatletter
\providecommand \@ifxundefined [1]{%
 \@ifx{#1\undefined}
}%
\providecommand \@ifnum [1]{%
 \ifnum #1\expandafter \@firstoftwo
 \else \expandafter \@secondoftwo
 \fi
}%
\providecommand \@ifx [1]{%
 \ifx #1\expandafter \@firstoftwo
 \else \expandafter \@secondoftwo
 \fi
}%
\providecommand \natexlab [1]{#1}%
\providecommand \enquote  [1]{``#1''}%
\providecommand \bibnamefont  [1]{#1}%
\providecommand \bibfnamefont [1]{#1}%
\providecommand \citenamefont [1]{#1}%
\providecommand \href@noop [0]{\@secondoftwo}%
\providecommand \href [0]{\begingroup \@sanitize@url \@href}%
\providecommand \@href[1]{\@@startlink{#1}\@@href}%
\providecommand \@@href[1]{\endgroup#1\@@endlink}%
\providecommand \@sanitize@url [0]{\catcode `\\12\catcode `\$12\catcode
  `\&12\catcode `\#12\catcode `\^12\catcode `\_12\catcode `\%12\relax}%
\providecommand \@@startlink[1]{}%
\providecommand \@@endlink[0]{}%
\providecommand \url  [0]{\begingroup\@sanitize@url \@url }%
\providecommand \@url [1]{\endgroup\@href {#1}{\urlprefix }}%
\providecommand \urlprefix  [0]{URL }%
\providecommand \Eprint [0]{\href }%
\providecommand \doibase [0]{https://doi.org/}%
\providecommand \selectlanguage [0]{\@gobble}%
\providecommand \bibinfo  [0]{\@secondoftwo}%
\providecommand \bibfield  [0]{\@secondoftwo}%
\providecommand \translation [1]{[#1]}%
\providecommand \BibitemOpen [0]{}%
\providecommand \bibitemStop [0]{}%
\providecommand \bibitemNoStop [0]{.\EOS\space}%
\providecommand \EOS [0]{\spacefactor3000\relax}%
\providecommand \BibitemShut  [1]{\csname bibitem#1\endcsname}%
\let\auto@bib@innerbib\@empty
\bibitem [{\citenamefont {Milo}\ and\ \citenamefont
  {Phillips}(2015)}]{milo2015cell}%
  \BibitemOpen
  \bibfield  {author} {\bibinfo {author} {\bibfnamefont {R.}~\bibnamefont
  {Milo}}\ and\ \bibinfo {author} {\bibfnamefont {R.}~\bibnamefont
  {Phillips}},\ }\href@noop {} {\emph {\bibinfo {title} {Cell biology by the
  numbers}}}\ (\bibinfo  {publisher} {Garland Science},\ \bibinfo {year}
  {2015})\BibitemShut {NoStop}%
\bibitem [{\citenamefont {No{\'e}}\ and\ \citenamefont
  {Clementi}(2017)}]{noe2017collective}%
  \BibitemOpen
  \bibfield  {author} {\bibinfo {author} {\bibfnamefont {F.}~\bibnamefont
  {No{\'e}}}\ and\ \bibinfo {author} {\bibfnamefont {C.}~\bibnamefont
  {Clementi}},\ }\bibfield  {title} {\bibinfo {title} {Collective variables for
  the study of long-time kinetics from molecular trajectories: theory and
  methods},\ }\href@noop {} {\bibfield  {journal} {\bibinfo  {journal} {Current
  opinion in structural biology}\ }\textbf {\bibinfo {volume} {43}},\ \bibinfo
  {pages} {141} (\bibinfo {year} {2017})}\BibitemShut {NoStop}%
\bibitem [{\citenamefont {Stephens}\ \emph {et~al.}(2008)\citenamefont
  {Stephens}, \citenamefont {Johnson-Kerner}, \citenamefont {Bialek},\ and\
  \citenamefont {Ryu}}]{stephens2008dimensionality}%
  \BibitemOpen
  \bibfield  {author} {\bibinfo {author} {\bibfnamefont {G.~J.}\ \bibnamefont
  {Stephens}}, \bibinfo {author} {\bibfnamefont {B.}~\bibnamefont
  {Johnson-Kerner}}, \bibinfo {author} {\bibfnamefont {W.}~\bibnamefont
  {Bialek}},\ and\ \bibinfo {author} {\bibfnamefont {W.~S.}\ \bibnamefont
  {Ryu}},\ }\bibfield  {title} {\bibinfo {title} {Dimensionality and dynamics
  in the behavior of c. elegans},\ }\href@noop {} {\bibfield  {journal}
  {\bibinfo  {journal} {PLoS computational biology}\ }\textbf {\bibinfo
  {volume} {4}},\ \bibinfo {pages} {e1000028} (\bibinfo {year}
  {2008})}\BibitemShut {NoStop}%
\bibitem [{\citenamefont {Sha}\ \emph {et~al.}(2003)\citenamefont {Sha},
  \citenamefont {Moore}, \citenamefont {Chen}, \citenamefont {Lassaletta},
  \citenamefont {Yi}, \citenamefont {Tyson},\ and\ \citenamefont
  {Sible}}]{sha2003hysteresis}%
  \BibitemOpen
  \bibfield  {author} {\bibinfo {author} {\bibfnamefont {W.}~\bibnamefont
  {Sha}}, \bibinfo {author} {\bibfnamefont {J.}~\bibnamefont {Moore}}, \bibinfo
  {author} {\bibfnamefont {K.}~\bibnamefont {Chen}}, \bibinfo {author}
  {\bibfnamefont {A.~D.}\ \bibnamefont {Lassaletta}}, \bibinfo {author}
  {\bibfnamefont {C.-S.}\ \bibnamefont {Yi}}, \bibinfo {author} {\bibfnamefont
  {J.~J.}\ \bibnamefont {Tyson}},\ and\ \bibinfo {author} {\bibfnamefont
  {J.~C.}\ \bibnamefont {Sible}},\ }\bibfield  {title} {\bibinfo {title}
  {Hysteresis drives cell-cycle transitions in xenopus laevis egg extracts},\
  }\href@noop {} {\bibfield  {journal} {\bibinfo  {journal} {Proceedings of the
  National Academy of Sciences}\ }\textbf {\bibinfo {volume} {100}},\ \bibinfo
  {pages} {975} (\bibinfo {year} {2003})}\BibitemShut {NoStop}%
\bibitem [{\citenamefont {Pomerening}\ \emph {et~al.}(2003)\citenamefont
  {Pomerening}, \citenamefont {Sontag},\ and\ \citenamefont
  {Ferrell~Jr}}]{pomerening2003building}%
  \BibitemOpen
  \bibfield  {author} {\bibinfo {author} {\bibfnamefont {J.~R.}\ \bibnamefont
  {Pomerening}}, \bibinfo {author} {\bibfnamefont {E.~D.}\ \bibnamefont
  {Sontag}},\ and\ \bibinfo {author} {\bibfnamefont {J.~E.}\ \bibnamefont
  {Ferrell~Jr}},\ }\bibfield  {title} {\bibinfo {title} {Building a cell cycle
  oscillator: hysteresis and bistability in the activation of cdc2},\
  }\href@noop {} {\bibfield  {journal} {\bibinfo  {journal} {Nature cell
  biology}\ }\textbf {\bibinfo {volume} {5}},\ \bibinfo {pages} {346} (\bibinfo
  {year} {2003})}\BibitemShut {NoStop}%
\bibitem [{\citenamefont {Coudreuse}\ and\ \citenamefont
  {Nurse}(2010)}]{coudreuse2010driving}%
  \BibitemOpen
  \bibfield  {author} {\bibinfo {author} {\bibfnamefont {D.}~\bibnamefont
  {Coudreuse}}\ and\ \bibinfo {author} {\bibfnamefont {P.}~\bibnamefont
  {Nurse}},\ }\bibfield  {title} {\bibinfo {title} {Driving the cell cycle with
  a minimal cdk control network},\ }\href@noop {} {\bibfield  {journal}
  {\bibinfo  {journal} {Nature}\ }\textbf {\bibinfo {volume} {468}},\ \bibinfo
  {pages} {1074} (\bibinfo {year} {2010})}\BibitemShut {NoStop}%
\bibitem [{\citenamefont {Rata}\ \emph {et~al.}(2018)\citenamefont {Rata},
  \citenamefont {Rodriguez}, \citenamefont {Joseph}, \citenamefont {Peter},
  \citenamefont {Iturra}, \citenamefont {Yang}, \citenamefont {Madzvamuse},
  \citenamefont {Ruppert}, \citenamefont {Samejima}, \citenamefont {Platani}
  \emph {et~al.}}]{rata2018two}%
  \BibitemOpen
  \bibfield  {author} {\bibinfo {author} {\bibfnamefont {S.}~\bibnamefont
  {Rata}}, \bibinfo {author} {\bibfnamefont {M.~F. S.~P.}\ \bibnamefont
  {Rodriguez}}, \bibinfo {author} {\bibfnamefont {S.}~\bibnamefont {Joseph}},
  \bibinfo {author} {\bibfnamefont {N.}~\bibnamefont {Peter}}, \bibinfo
  {author} {\bibfnamefont {F.~E.}\ \bibnamefont {Iturra}}, \bibinfo {author}
  {\bibfnamefont {F.}~\bibnamefont {Yang}}, \bibinfo {author} {\bibfnamefont
  {A.}~\bibnamefont {Madzvamuse}}, \bibinfo {author} {\bibfnamefont {J.~G.}\
  \bibnamefont {Ruppert}}, \bibinfo {author} {\bibfnamefont {K.}~\bibnamefont
  {Samejima}}, \bibinfo {author} {\bibfnamefont {M.}~\bibnamefont {Platani}},
  \emph {et~al.},\ }\bibfield  {title} {\bibinfo {title} {Two interlinked
  bistable switches govern mitotic control in mammalian cells},\ }\href@noop {}
  {\bibfield  {journal} {\bibinfo  {journal} {Current biology}\ }\textbf
  {\bibinfo {volume} {28}},\ \bibinfo {pages} {3824} (\bibinfo {year}
  {2018})}\BibitemShut {NoStop}%
\bibitem [{\citenamefont {Schwarz}\ \emph {et~al.}(2018)\citenamefont
  {Schwarz}, \citenamefont {Johnson}, \citenamefont {K{\~o}ivom{\"a}gi},
  \citenamefont {Zatulovskiy}, \citenamefont {Kravitz}, \citenamefont
  {Doncic},\ and\ \citenamefont {Skotheim}}]{schwarz2018precise}%
  \BibitemOpen
  \bibfield  {author} {\bibinfo {author} {\bibfnamefont {C.}~\bibnamefont
  {Schwarz}}, \bibinfo {author} {\bibfnamefont {A.}~\bibnamefont {Johnson}},
  \bibinfo {author} {\bibfnamefont {M.}~\bibnamefont {K{\~o}ivom{\"a}gi}},
  \bibinfo {author} {\bibfnamefont {E.}~\bibnamefont {Zatulovskiy}}, \bibinfo
  {author} {\bibfnamefont {C.~J.}\ \bibnamefont {Kravitz}}, \bibinfo {author}
  {\bibfnamefont {A.}~\bibnamefont {Doncic}},\ and\ \bibinfo {author}
  {\bibfnamefont {J.~M.}\ \bibnamefont {Skotheim}},\ }\bibfield  {title}
  {\bibinfo {title} {A precise cdk activity threshold determines passage
  through the restriction point},\ }\href@noop {} {\bibfield  {journal}
  {\bibinfo  {journal} {Molecular cell}\ }\textbf {\bibinfo {volume} {69}},\
  \bibinfo {pages} {253} (\bibinfo {year} {2018})}\BibitemShut {NoStop}%
\bibitem [{\citenamefont {Domingo-Sananes}\ \emph {et~al.}(2011)\citenamefont
  {Domingo-Sananes}, \citenamefont {Kapuy}, \citenamefont {Hunt},\ and\
  \citenamefont {Novak}}]{domingo2011switches}%
  \BibitemOpen
  \bibfield  {author} {\bibinfo {author} {\bibfnamefont {M.~R.}\ \bibnamefont
  {Domingo-Sananes}}, \bibinfo {author} {\bibfnamefont {O.}~\bibnamefont
  {Kapuy}}, \bibinfo {author} {\bibfnamefont {T.}~\bibnamefont {Hunt}},\ and\
  \bibinfo {author} {\bibfnamefont {B.}~\bibnamefont {Novak}},\ }\bibfield
  {title} {\bibinfo {title} {Switches and latches: a biochemical tug-of-war
  between the kinases and phosphatases that control mitosis},\ }\href@noop {}
  {\bibfield  {journal} {\bibinfo  {journal} {Philosophical Transactions of the
  Royal Society B: Biological Sciences}\ }\textbf {\bibinfo {volume} {366}},\
  \bibinfo {pages} {3584} (\bibinfo {year} {2011})}\BibitemShut {NoStop}%
\bibitem [{\citenamefont {Nachman}\ \emph {et~al.}(2007)\citenamefont
  {Nachman}, \citenamefont {Regev},\ and\ \citenamefont
  {Ramanathan}}]{nachman2007dissecting}%
  \BibitemOpen
  \bibfield  {author} {\bibinfo {author} {\bibfnamefont {I.}~\bibnamefont
  {Nachman}}, \bibinfo {author} {\bibfnamefont {A.}~\bibnamefont {Regev}},\
  and\ \bibinfo {author} {\bibfnamefont {S.}~\bibnamefont {Ramanathan}},\
  }\bibfield  {title} {\bibinfo {title} {Dissecting timing variability in yeast
  meiosis},\ }\href@noop {} {\bibfield  {journal} {\bibinfo  {journal} {Cell}\
  }\textbf {\bibinfo {volume} {131}},\ \bibinfo {pages} {544} (\bibinfo {year}
  {2007})}\BibitemShut {NoStop}%
\bibitem [{\citenamefont {Pardee}(1974)}]{pardee1974restriction}%
  \BibitemOpen
  \bibfield  {author} {\bibinfo {author} {\bibfnamefont {A.~B.}\ \bibnamefont
  {Pardee}},\ }\bibfield  {title} {\bibinfo {title} {A restriction point for
  control of normal animal cell proliferation},\ }\href@noop {} {\bibfield
  {journal} {\bibinfo  {journal} {Proceedings of the National Academy of
  Sciences}\ }\textbf {\bibinfo {volume} {71}},\ \bibinfo {pages} {1286}
  (\bibinfo {year} {1974})}\BibitemShut {NoStop}%
\bibitem [{\citenamefont {Ahrends}\ \emph {et~al.}(2014)\citenamefont
  {Ahrends}, \citenamefont {Ota}, \citenamefont {Kovary}, \citenamefont {Kudo},
  \citenamefont {Park},\ and\ \citenamefont {Teruel}}]{ahrends2014controlling}%
  \BibitemOpen
  \bibfield  {author} {\bibinfo {author} {\bibfnamefont {R.}~\bibnamefont
  {Ahrends}}, \bibinfo {author} {\bibfnamefont {A.}~\bibnamefont {Ota}},
  \bibinfo {author} {\bibfnamefont {K.~M.}\ \bibnamefont {Kovary}}, \bibinfo
  {author} {\bibfnamefont {T.}~\bibnamefont {Kudo}}, \bibinfo {author}
  {\bibfnamefont {B.~O.}\ \bibnamefont {Park}},\ and\ \bibinfo {author}
  {\bibfnamefont {M.~N.}\ \bibnamefont {Teruel}},\ }\bibfield  {title}
  {\bibinfo {title} {Controlling low rates of cell differentiation through
  noise and ultrahigh feedback},\ }\href@noop {} {\bibfield  {journal}
  {\bibinfo  {journal} {Science}\ }\textbf {\bibinfo {volume} {344}},\ \bibinfo
  {pages} {1384} (\bibinfo {year} {2014})}\BibitemShut {NoStop}%
\bibitem [{\citenamefont {Maamar}\ \emph {et~al.}(2007)\citenamefont {Maamar},
  \citenamefont {Raj},\ and\ \citenamefont {Dubnau}}]{maamar2007noise}%
  \BibitemOpen
  \bibfield  {author} {\bibinfo {author} {\bibfnamefont {H.}~\bibnamefont
  {Maamar}}, \bibinfo {author} {\bibfnamefont {A.}~\bibnamefont {Raj}},\ and\
  \bibinfo {author} {\bibfnamefont {D.}~\bibnamefont {Dubnau}},\ }\bibfield
  {title} {\bibinfo {title} {Noise in gene expression determines cell fate in
  bacillus subtilis},\ }\href@noop {} {\bibfield  {journal} {\bibinfo
  {journal} {Science}\ }\textbf {\bibinfo {volume} {317}},\ \bibinfo {pages}
  {526} (\bibinfo {year} {2007})}\BibitemShut {NoStop}%
\bibitem [{\citenamefont {Xiong}\ and\ \citenamefont
  {Ferrell~Jr}(2003)}]{xiong2003positive}%
  \BibitemOpen
  \bibfield  {author} {\bibinfo {author} {\bibfnamefont {W.}~\bibnamefont
  {Xiong}}\ and\ \bibinfo {author} {\bibfnamefont {J.~E.}\ \bibnamefont
  {Ferrell~Jr}},\ }\bibfield  {title} {\bibinfo {title} {A
  positive-feedback-based bistable ‘memory module’that governs a cell fate
  decision},\ }\href@noop {} {\bibfield  {journal} {\bibinfo  {journal}
  {Nature}\ }\textbf {\bibinfo {volume} {426}},\ \bibinfo {pages} {460}
  (\bibinfo {year} {2003})}\BibitemShut {NoStop}%
\bibitem [{\citenamefont {Losick}\ and\ \citenamefont
  {Desplan}(2008)}]{losick2008stochasticity}%
  \BibitemOpen
  \bibfield  {author} {\bibinfo {author} {\bibfnamefont {R.}~\bibnamefont
  {Losick}}\ and\ \bibinfo {author} {\bibfnamefont {C.}~\bibnamefont
  {Desplan}},\ }\bibfield  {title} {\bibinfo {title} {Stochasticity and cell
  fate},\ }\href@noop {} {\bibfield  {journal} {\bibinfo  {journal} {science}\
  }\textbf {\bibinfo {volume} {320}},\ \bibinfo {pages} {65} (\bibinfo {year}
  {2008})}\BibitemShut {NoStop}%
\bibitem [{\citenamefont {Bal{\'a}zsi}\ \emph {et~al.}(2011)\citenamefont
  {Bal{\'a}zsi}, \citenamefont {van Oudenaarden},\ and\ \citenamefont
  {Collins}}]{balazsi2011cellular}%
  \BibitemOpen
  \bibfield  {author} {\bibinfo {author} {\bibfnamefont {G.}~\bibnamefont
  {Bal{\'a}zsi}}, \bibinfo {author} {\bibfnamefont {A.}~\bibnamefont {van
  Oudenaarden}},\ and\ \bibinfo {author} {\bibfnamefont {J.~J.}\ \bibnamefont
  {Collins}},\ }\bibfield  {title} {\bibinfo {title} {Cellular decision making
  and biological noise: from microbes to mammals},\ }\href@noop {} {\bibfield
  {journal} {\bibinfo  {journal} {Cell}\ }\textbf {\bibinfo {volume} {144}},\
  \bibinfo {pages} {910} (\bibinfo {year} {2011})}\BibitemShut {NoStop}%
\bibitem [{\citenamefont {Hanes}\ and\ \citenamefont
  {Schall}(1996)}]{hanes1996neural}%
  \BibitemOpen
  \bibfield  {author} {\bibinfo {author} {\bibfnamefont {D.~P.}\ \bibnamefont
  {Hanes}}\ and\ \bibinfo {author} {\bibfnamefont {J.~D.}\ \bibnamefont
  {Schall}},\ }\bibfield  {title} {\bibinfo {title} {Neural control of
  voluntary movement initiation},\ }\href@noop {} {\bibfield  {journal}
  {\bibinfo  {journal} {Science}\ }\textbf {\bibinfo {volume} {274}},\ \bibinfo
  {pages} {427} (\bibinfo {year} {1996})}\BibitemShut {NoStop}%
\bibitem [{\citenamefont {Hanks}\ \emph {et~al.}(2015)\citenamefont {Hanks},
  \citenamefont {Kopec}, \citenamefont {Brunton}, \citenamefont {Duan},
  \citenamefont {Erlich},\ and\ \citenamefont {Brody}}]{hanks2015distinct}%
  \BibitemOpen
  \bibfield  {author} {\bibinfo {author} {\bibfnamefont {T.~D.}\ \bibnamefont
  {Hanks}}, \bibinfo {author} {\bibfnamefont {C.~D.}\ \bibnamefont {Kopec}},
  \bibinfo {author} {\bibfnamefont {B.~W.}\ \bibnamefont {Brunton}}, \bibinfo
  {author} {\bibfnamefont {C.~A.}\ \bibnamefont {Duan}}, \bibinfo {author}
  {\bibfnamefont {J.~C.}\ \bibnamefont {Erlich}},\ and\ \bibinfo {author}
  {\bibfnamefont {C.~D.}\ \bibnamefont {Brody}},\ }\bibfield  {title} {\bibinfo
  {title} {Distinct relationships of parietal and prefrontal cortices to
  evidence accumulation},\ }\href@noop {} {\bibfield  {journal} {\bibinfo
  {journal} {Nature}\ }\textbf {\bibinfo {volume} {520}},\ \bibinfo {pages}
  {220} (\bibinfo {year} {2015})}\BibitemShut {NoStop}%
\bibitem [{\citenamefont {Ratcliff}\ \emph {et~al.}(2016)\citenamefont
  {Ratcliff}, \citenamefont {Smith}, \citenamefont {Brown},\ and\ \citenamefont
  {McKoon}}]{ratcliff2016diffusion}%
  \BibitemOpen
  \bibfield  {author} {\bibinfo {author} {\bibfnamefont {R.}~\bibnamefont
  {Ratcliff}}, \bibinfo {author} {\bibfnamefont {P.~L.}\ \bibnamefont {Smith}},
  \bibinfo {author} {\bibfnamefont {S.~D.}\ \bibnamefont {Brown}},\ and\
  \bibinfo {author} {\bibfnamefont {G.}~\bibnamefont {McKoon}},\ }\bibfield
  {title} {\bibinfo {title} {Diffusion decision model: current issues and
  history},\ }\href@noop {} {\bibfield  {journal} {\bibinfo  {journal} {Trends
  in cognitive sciences}\ }\textbf {\bibinfo {volume} {20}},\ \bibinfo {pages}
  {260} (\bibinfo {year} {2016})}\BibitemShut {NoStop}%
\bibitem [{\citenamefont {Ratcliff}\ and\ \citenamefont
  {McKoon}(2008)}]{ratcliff2008diffusion}%
  \BibitemOpen
  \bibfield  {author} {\bibinfo {author} {\bibfnamefont {R.}~\bibnamefont
  {Ratcliff}}\ and\ \bibinfo {author} {\bibfnamefont {G.}~\bibnamefont
  {McKoon}},\ }\bibfield  {title} {\bibinfo {title} {The diffusion decision
  model: theory and data for two-choice decision tasks},\ }\href@noop {}
  {\bibfield  {journal} {\bibinfo  {journal} {Neural computation}\ }\textbf
  {\bibinfo {volume} {20}},\ \bibinfo {pages} {873} (\bibinfo {year}
  {2008})}\BibitemShut {NoStop}%
\bibitem [{\citenamefont {Brunton}\ \emph {et~al.}(2013)\citenamefont
  {Brunton}, \citenamefont {Botvinick},\ and\ \citenamefont
  {Brody}}]{brunton2013rats}%
  \BibitemOpen
  \bibfield  {author} {\bibinfo {author} {\bibfnamefont {B.~W.}\ \bibnamefont
  {Brunton}}, \bibinfo {author} {\bibfnamefont {M.~M.}\ \bibnamefont
  {Botvinick}},\ and\ \bibinfo {author} {\bibfnamefont {C.~D.}\ \bibnamefont
  {Brody}},\ }\bibfield  {title} {\bibinfo {title} {Rats and humans can
  optimally accumulate evidence for decision-making},\ }\href@noop {}
  {\bibfield  {journal} {\bibinfo  {journal} {Science}\ }\textbf {\bibinfo
  {volume} {340}},\ \bibinfo {pages} {95} (\bibinfo {year} {2013})}\BibitemShut
  {NoStop}%
\bibitem [{\citenamefont {Churchland}\ \emph {et~al.}(2011)\citenamefont
  {Churchland}, \citenamefont {Kiani}, \citenamefont {Chaudhuri}, \citenamefont
  {Wang}, \citenamefont {Pouget},\ and\ \citenamefont
  {Shadlen}}]{churchland2011variance}%
  \BibitemOpen
  \bibfield  {author} {\bibinfo {author} {\bibfnamefont {A.~K.}\ \bibnamefont
  {Churchland}}, \bibinfo {author} {\bibfnamefont {R.}~\bibnamefont {Kiani}},
  \bibinfo {author} {\bibfnamefont {R.}~\bibnamefont {Chaudhuri}}, \bibinfo
  {author} {\bibfnamefont {X.-J.}\ \bibnamefont {Wang}}, \bibinfo {author}
  {\bibfnamefont {A.}~\bibnamefont {Pouget}},\ and\ \bibinfo {author}
  {\bibfnamefont {M.~N.}\ \bibnamefont {Shadlen}},\ }\bibfield  {title}
  {\bibinfo {title} {Variance as a signature of neural computations during
  decision making},\ }\href@noop {} {\bibfield  {journal} {\bibinfo  {journal}
  {Neuron}\ }\textbf {\bibinfo {volume} {69}},\ \bibinfo {pages} {818}
  (\bibinfo {year} {2011})}\BibitemShut {NoStop}%
\bibitem [{\citenamefont {Roitman}\ and\ \citenamefont
  {Shadlen}(2002)}]{roitman2002response}%
  \BibitemOpen
  \bibfield  {author} {\bibinfo {author} {\bibfnamefont {J.~D.}\ \bibnamefont
  {Roitman}}\ and\ \bibinfo {author} {\bibfnamefont {M.~N.}\ \bibnamefont
  {Shadlen}},\ }\bibfield  {title} {\bibinfo {title} {Response of neurons in
  the lateral intraparietal area during a combined visual discrimination
  reaction time task},\ }\href@noop {} {\bibfield  {journal} {\bibinfo
  {journal} {Journal of neuroscience}\ }\textbf {\bibinfo {volume} {22}},\
  \bibinfo {pages} {9475} (\bibinfo {year} {2002})}\BibitemShut {NoStop}%
\bibitem [{\citenamefont {Altschuler}\ and\ \citenamefont
  {Wu}(2010)}]{altschuler2010cellular}%
  \BibitemOpen
  \bibfield  {author} {\bibinfo {author} {\bibfnamefont {S.~J.}\ \bibnamefont
  {Altschuler}}\ and\ \bibinfo {author} {\bibfnamefont {L.~F.}\ \bibnamefont
  {Wu}},\ }\bibfield  {title} {\bibinfo {title} {Cellular heterogeneity: do
  differences make a difference?},\ }\href@noop {} {\bibfield  {journal}
  {\bibinfo  {journal} {Cell}\ }\textbf {\bibinfo {volume} {141}},\ \bibinfo
  {pages} {559} (\bibinfo {year} {2010})}\BibitemShut {NoStop}%
\bibitem [{\citenamefont {Komin}\ and\ \citenamefont
  {Skupin}(2017)}]{komin2017address}%
  \BibitemOpen
  \bibfield  {author} {\bibinfo {author} {\bibfnamefont {N.}~\bibnamefont
  {Komin}}\ and\ \bibinfo {author} {\bibfnamefont {A.}~\bibnamefont {Skupin}},\
  }\bibfield  {title} {\bibinfo {title} {How to address cellular heterogeneity
  by distribution biology},\ }\href@noop {} {\bibfield  {journal} {\bibinfo
  {journal} {Current Opinion in Systems Biology}\ }\textbf {\bibinfo {volume}
  {3}},\ \bibinfo {pages} {154} (\bibinfo {year} {2017})}\BibitemShut {NoStop}%
\bibitem [{\citenamefont {Gough}\ \emph {et~al.}(2017)\citenamefont {Gough},
  \citenamefont {Stern}, \citenamefont {Maier}, \citenamefont {Lezon},
  \citenamefont {Shun}, \citenamefont {Chennubhotla}, \citenamefont {Schurdak},
  \citenamefont {Haney},\ and\ \citenamefont {Taylor}}]{gough2017biologically}%
  \BibitemOpen
  \bibfield  {author} {\bibinfo {author} {\bibfnamefont {A.}~\bibnamefont
  {Gough}}, \bibinfo {author} {\bibfnamefont {A.~M.}\ \bibnamefont {Stern}},
  \bibinfo {author} {\bibfnamefont {J.}~\bibnamefont {Maier}}, \bibinfo
  {author} {\bibfnamefont {T.}~\bibnamefont {Lezon}}, \bibinfo {author}
  {\bibfnamefont {T.-Y.}\ \bibnamefont {Shun}}, \bibinfo {author}
  {\bibfnamefont {C.}~\bibnamefont {Chennubhotla}}, \bibinfo {author}
  {\bibfnamefont {M.~E.}\ \bibnamefont {Schurdak}}, \bibinfo {author}
  {\bibfnamefont {S.~A.}\ \bibnamefont {Haney}},\ and\ \bibinfo {author}
  {\bibfnamefont {D.~L.}\ \bibnamefont {Taylor}},\ }\bibfield  {title}
  {\bibinfo {title} {Biologically relevant heterogeneity: metrics and practical
  insights},\ }\href@noop {} {\bibfield  {journal} {\bibinfo  {journal} {Slas
  Discovery: Advancing Life Sciences R\&D}\ }\textbf {\bibinfo {volume} {22}},\
  \bibinfo {pages} {213} (\bibinfo {year} {2017})}\BibitemShut {NoStop}%
\bibitem [{\citenamefont {Lenner}\ \emph
  {et~al.}(2023{\natexlab{a}})\citenamefont {Lenner}, \citenamefont {Eule},
  \citenamefont {Gro{\ss}hans},\ and\ \citenamefont
  {Wolf}}]{lenner2023reverse1}%
  \BibitemOpen
  \bibfield  {author} {\bibinfo {author} {\bibfnamefont {N.}~\bibnamefont
  {Lenner}}, \bibinfo {author} {\bibfnamefont {S.}~\bibnamefont {Eule}},
  \bibinfo {author} {\bibfnamefont {J.}~\bibnamefont {Gro{\ss}hans}},\ and\
  \bibinfo {author} {\bibfnamefont {F.}~\bibnamefont {Wolf}},\ }\bibfield
  {title} {\bibinfo {title} {Reverse-time analysis uncovers universality
  classes in directional biological dynamics},\ }\href@noop {} {\bibfield
  {journal} {\bibinfo  {journal} {arXiv preprint arXiv:2304.03226}\ } (\bibinfo
  {year} {2023}{\natexlab{a}})}\BibitemShut {NoStop}%
\bibitem [{\citenamefont {Lenner}\ \emph
  {et~al.}(2023{\natexlab{b}})\citenamefont {Lenner}, \citenamefont
  {H{\"a}ring}, \citenamefont {Eule}, \citenamefont {Gro{\ss}hans},\ and\
  \citenamefont {Wolf}}]{lenner2023reverse2}%
  \BibitemOpen
  \bibfield  {author} {\bibinfo {author} {\bibfnamefont {N.}~\bibnamefont
  {Lenner}}, \bibinfo {author} {\bibfnamefont {M.}~\bibnamefont {H{\"a}ring}},
  \bibinfo {author} {\bibfnamefont {S.}~\bibnamefont {Eule}}, \bibinfo {author}
  {\bibfnamefont {J.}~\bibnamefont {Gro{\ss}hans}},\ and\ \bibinfo {author}
  {\bibfnamefont {F.}~\bibnamefont {Wolf}},\ }\bibfield  {title} {\bibinfo
  {title} {Reverse-time analysis and boundary classification of directional
  biological dynamics with multiplicative noise},\ }\href@noop {} {\bibfield
  {journal} {\bibinfo  {journal} {arXiv preprint arXiv:2304.04279}\ } (\bibinfo
  {year} {2023}{\natexlab{b}})}\BibitemShut {NoStop}%
\bibitem [{\citenamefont {Gillespie}(2000)}]{gillespie2000chemical}%
  \BibitemOpen
  \bibfield  {author} {\bibinfo {author} {\bibfnamefont {D.~T.}\ \bibnamefont
  {Gillespie}},\ }\bibfield  {title} {\bibinfo {title} {The chemical langevin
  equation},\ }\href@noop {} {\bibfield  {journal} {\bibinfo  {journal} {The
  Journal of Chemical Physics}\ }\textbf {\bibinfo {volume} {113}},\ \bibinfo
  {pages} {297} (\bibinfo {year} {2000})}\BibitemShut {NoStop}%
\bibitem [{\citenamefont {Berezhkovskii}\ and\ \citenamefont
  {Szabo}(2011)}]{berezhkovskii2011time}%
  \BibitemOpen
  \bibfield  {author} {\bibinfo {author} {\bibfnamefont {A.}~\bibnamefont
  {Berezhkovskii}}\ and\ \bibinfo {author} {\bibfnamefont {A.}~\bibnamefont
  {Szabo}},\ }\bibfield  {title} {\bibinfo {title} {Time scale separation leads
  to position-dependent diffusion along a slow coordinate},\ }\href@noop {}
  {\bibfield  {journal} {\bibinfo  {journal} {The Journal of chemical physics}\
  }\textbf {\bibinfo {volume} {135}},\ \bibinfo {pages} {074108} (\bibinfo
  {year} {2011})}\BibitemShut {NoStop}%
\bibitem [{\citenamefont {Swain}\ \emph {et~al.}(2002)\citenamefont {Swain},
  \citenamefont {Elowitz},\ and\ \citenamefont {Siggia}}]{swain2002intrinsic}%
  \BibitemOpen
  \bibfield  {author} {\bibinfo {author} {\bibfnamefont {P.~S.}\ \bibnamefont
  {Swain}}, \bibinfo {author} {\bibfnamefont {M.~B.}\ \bibnamefont {Elowitz}},\
  and\ \bibinfo {author} {\bibfnamefont {E.~D.}\ \bibnamefont {Siggia}},\
  }\bibfield  {title} {\bibinfo {title} {Intrinsic and extrinsic contributions
  to stochasticity in gene expression},\ }\href@noop {} {\bibfield  {journal}
  {\bibinfo  {journal} {Proceedings of the National Academy of Sciences}\
  }\textbf {\bibinfo {volume} {99}},\ \bibinfo {pages} {12795} (\bibinfo {year}
  {2002})}\BibitemShut {NoStop}%
\bibitem [{\citenamefont {Van~Kampen}(1981)}]{van1981ito}%
  \BibitemOpen
  \bibfield  {author} {\bibinfo {author} {\bibfnamefont {N.}~\bibnamefont
  {Van~Kampen}},\ }\bibfield  {title} {\bibinfo {title} {It{\^o} versus
  stratonovich},\ }\href@noop {} {\bibfield  {journal} {\bibinfo  {journal}
  {Journal of Statistical Physics}\ }\textbf {\bibinfo {volume} {24}},\
  \bibinfo {pages} {175} (\bibinfo {year} {1981})}\BibitemShut {NoStop}%
\end{thebibliography}%


\providecommand{\noopsort}[1]{}\providecommand{\singleletter}[1]{#1}%
\begin{thebibliography}{3}%
\makeatletter
\providecommand \@ifxundefined [1]{%
 \@ifx{#1\undefined}
}%
\providecommand \@ifnum [1]{%
 \ifnum #1\expandafter \@firstoftwo
 \else \expandafter \@secondoftwo
 \fi
}%
\providecommand \@ifx [1]{%
 \ifx #1\expandafter \@firstoftwo
 \else \expandafter \@secondoftwo
 \fi
}%
\providecommand \natexlab [1]{#1}%
\providecommand \enquote  [1]{``#1''}%
\providecommand \bibnamefont  [1]{#1}%
\providecommand \bibfnamefont [1]{#1}%
\providecommand \citenamefont [1]{#1}%
\providecommand \href@noop [0]{\@secondoftwo}%
\providecommand \href [0]{\begingroup \@sanitize@url \@href}%
\providecommand \@href[1]{\@@startlink{#1}\@@href}%
\providecommand \@@href[1]{\endgroup#1\@@endlink}%
\providecommand \@sanitize@url [0]{\catcode `\\12\catcode `\$12\catcode
  `\&12\catcode `\#12\catcode `\^12\catcode `\_12\catcode `\%12\relax}%
\providecommand \@@startlink[1]{}%
\providecommand \@@endlink[0]{}%
\providecommand \url  [0]{\begingroup\@sanitize@url \@url }%
\providecommand \@url [1]{\endgroup\@href {#1}{\urlprefix }}%
\providecommand \urlprefix  [0]{URL }%
\providecommand \Eprint [0]{\href }%
\providecommand \doibase [0]{https://doi.org/}%
\providecommand \selectlanguage [0]{\@gobble}%
\providecommand \bibinfo  [0]{\@secondoftwo}%
\providecommand \bibfield  [0]{\@secondoftwo}%
\providecommand \translation [1]{[#1]}%
\providecommand \BibitemOpen [0]{}%
\providecommand \bibitemStop [0]{}%
\providecommand \bibitemNoStop [0]{.\EOS\space}%
\providecommand \EOS [0]{\spacefactor3000\relax}%
\providecommand \BibitemShut  [1]{\csname bibitem#1\endcsname}%
\let\auto@bib@innerbib\@empty
\bibitem [{\citenamefont {Lenner}\ \emph {et~al.}(2023)\citenamefont {Lenner},
  \citenamefont {H{\"a}ring}, \citenamefont {Eule}, \citenamefont
  {Gro{\ss}hans},\ and\ \citenamefont {Wolf}}]{lenner2023reverse2}%
  \BibitemOpen
  \bibfield  {author} {\bibinfo {author} {\bibfnamefont {N.}~\bibnamefont
  {Lenner}}, \bibinfo {author} {\bibfnamefont {M.}~\bibnamefont {H{\"a}ring}},
  \bibinfo {author} {\bibfnamefont {S.}~\bibnamefont {Eule}}, \bibinfo {author}
  {\bibfnamefont {J.}~\bibnamefont {Gro{\ss}hans}},\ and\ \bibinfo {author}
  {\bibfnamefont {F.}~\bibnamefont {Wolf}},\ }\bibfield  {title} {\bibinfo
  {title} {Reverse-time analysis and boundary classification of directional
  biological dynamics with multiplicative noise},\ }\href@noop {} {\bibfield
  {journal} {\bibinfo  {journal} {arXiv preprint arXiv:2304.04279}\ } (\bibinfo
  {year} {2023})}\BibitemShut {NoStop}%
\bibitem [{\citenamefont {Gardiner}(1985)}]{gardiner1985handbook}%
  \BibitemOpen
  \bibfield  {author} {\bibinfo {author} {\bibfnamefont {C.~W.}\ \bibnamefont
  {Gardiner}},\ }\bibfield  {title} {\bibinfo {title} {Handbook of stochastic
  methods for physics, chemistry and the natural sciences, vol. 13 of},\
  }\href@noop {} {\bibfield  {journal} {\bibinfo  {journal} {Springer series in
  synergetics}\ } (\bibinfo {year} {1985})}\BibitemShut {NoStop}%
\bibitem [{\citenamefont {Bray}(2000)}]{bray2000random}%
  \BibitemOpen
  \bibfield  {author} {\bibinfo {author} {\bibfnamefont {A.}~\bibnamefont
  {Bray}},\ }\bibfield  {title} {\bibinfo {title} {Random walks in logarithmic
  and power-law potentials, nonuniversal persistence, and vortex dynamics in
  the two-dimensional xy model},\ }\href@noop {} {\bibfield  {journal}
  {\bibinfo  {journal} {Physical Review E}\ }\textbf {\bibinfo {volume} {62}},\
  \bibinfo {pages} {103} (\bibinfo {year} {2000})}\BibitemShut {NoStop}%
\end{thebibliography}%

\end{document}



\title{Supplementary Information - Signatures of heterogeneity in the statistical structure of target state aligned ensembles}

\author{Nicolas Lenner}
 \altaffiliation[Currently at ]{Simons Center for Systems Biology, School of Natural Sciences, Institute for Advanced Study, Princeton, New Jersey, USA.}
  \email{Lenner@ias.edu}
 \affiliation{Max Planck Institute for Dynamics and Self-Organization, Göttingen, Germany}

\author{Matthias Häring}
\affiliation{Max Planck Institute for Dynamics and Self-Organization, Göttingen, Germany}
\affiliation{Göttingen Campus Institute for Dynamics of Biological Networks, University of Göttingen, Göttingen, Germany}

\author{Stephan Eule}%
\affiliation{Max Planck Institute for Dynamics and Self-Organization, Göttingen, Germany}
\affiliation{German Primate Center—Leibniz Institute for Primate Research, Goettingen, Germany}

\author{Jörg Großhans}
\affiliation{Department of Biology, Philipps University Marburg, Marburg, Germany}
\affiliation{Göttingen Campus Institute for Dynamics of Biological Networks, University of Göttingen, Göttingen, Germany}

\author{Fred Wolf}
 \email{Fred.Wolf@ds.mpg.de}
\affiliation{Max Planck Institute for Dynamics and Self-Organization, Göttingen, Germany}
\affiliation{Göttingen Campus Institute for Dynamics of Biological Networks, University of Göttingen, Göttingen, Germany}
\affiliation{Max Planck Institute for Multidisciplinary Sciences, Göttingen, Germany}
\affiliation{Institute for the Dynamics of Complex Systems, University of Göttingen, Göttingen, Germany}
\affiliation{Center for Biostructural Imaging of Neurodegeneration, Göttingen, Germany}
\affiliation{Bernstein Center for Computational Neuroscience Göttingen, Göttingen, Germany}


\maketitle


%
%
%
%
%
%
%
%

\newpage
\tableofcontents
\newpage


\section{Small noise approximation of moments for force dominated homogeneous TSA ensembles}
\label{sec_hom_sN_exp}
In this section we reiterate our previous calculations for the small noise expansion for force dominated homogeneous TSA ensembles. We closely follow the version presented in Lenner et.~al.~\cite{lenner2023reverse2} which is based on the small noise expansion for Langevin equations discussed in Gardiner \cite{gardiner1985handbook}. These calculations provide the background for the modifications which we apply to adapt this approach to dynamics with parametric heterogeneity. 

All calculations start from the small $L$ expansion of the force dominated reverse-time Langevin equation
 \begin{align}
\label{sup_tr_lv_tsa_sn_multN_alpha_sm_beta_simp}
  dL(\tau) =
  \left(
    \gamma L^\alpha
      -  D (\alpha-\beta) L^{\beta-1} 
      +
      \mathcal{O} \left( \frac{D^2}{L^{2+\alpha-2\beta}} \right)
 \right)
       d\tau
+
      \sqrt{D \, L^\beta} \ d W_\tau
     \ .
\end{align}
which we derived in \cite{lenner2023reverse2} and restate in the main text. It is strictly valid for $\alpha < \beta-1$ with $\beta<2$ and $\alpha<1$. Its validity can be extended to also include the interval $\beta-1 \le \alpha < \beta$ if $D\to 0$ as $L\to 0$. 

Similarly, the small noise expressions we derive here, is strictly valid only for $\alpha < \beta-1$. The extension to $\alpha < \beta$ will nevertheless yield good approximations for most cases as the additional requirement of $D\to 0$ as $L\to 0$ is by construction, that is for a small $D$ approximation, sufficiently fulfilled.

We start our derivation of the small noise expansion with a rewritten version of Eq.~\eqref{sup_tr_lv_tsa_sn_multN_alpha_sm_beta_simp}
\begin{align}
  dL = a(L) d\tau + \epsilon^2 b(L) d\tau + \epsilon \; c(L)  d W_\tau
  \ ,
\end{align}
with  $\sqrt{D}$ substituted by the order parameter $\epsilon$. We next expand 
\begin{align}
\label{sup_epsilon_expansion}
 L(\tau) = L_0(\tau) + \epsilon L_1(\tau) + \epsilon^2 L_2(\tau) + ...
\end{align}
for small $\epsilon$ and around the deterministic solution $L_0(\tau)$. Similarly we expand $a(L)$, $b(c)$ and $c(L)$. Assuming that $a(L)$ can be written as
\begin{align}
 a(L) = a(L_0 + \epsilon L_1 + \epsilon^2 L_2 + ...)
      = a_0(L_0) + \epsilon a_1(L_0,L_1) + \epsilon^2 a_2(L_0,L_1,L_2)
      + ... \ ,
\end{align}
a general expansion reads
\begin{align}
 a(L) = a\left(L_0 + \sum_{m=1}^\infty \epsilon^m L_m \right)
 = \sum_{p=0}^\infty \frac{1}{p!} \frac{d^p a(L_0)}{d L_0} 
 \left(
 \sum_{p=0}^\infty \epsilon^m L_m
 \right)^p
 \ .
\end{align}
An analogous expression holds for $b(L)$ and $c(L)$.
After sorting terms we find for the first three terms
\begin{align}
 &a_0(L_0) = a(L_0) 
 \\
 &a_1(L_0,L_1) = L_1 \frac{d a(L_0)}{d L_0}
 \\
 \label{sup_a2}
 &a_2(L_0,L_1,L_2) = L_2 \frac{d a(L_0)}{d L_0} 
 + \frac{1}{2} L_1^2 \frac{d^2 a(L_0)}{d L_0^2}
 \ .
\end{align}
Following the same expansion scheme for $b(L)$ and $c(L)$ we arrive at an ordered set of stochastic differential equations
\begin{align}
\label{sup_dx0_mult}
 &d L_0 = a(L_0) d\tau
 \\
\label{sup_dx1_mult}
 &d L_1 = a_1(L_1,L_0) d\tau + c(L_0) d W_\tau
 \\
 \label{sup_dx2_mult}
 &d L_2 = a_2(L_2,L_1,L_0) d\tau + b(L_0) d\tau + c(L_1,L_0) dW_\tau
 \ ,
\end{align}
which we truncate after the second contributing order to the full solution $L(\tau)$. For $f(L)=-\gamma L^\alpha$ and $D(L)=D L^\beta$, as defined above, the terms evaluate as follows. The zeroth order term 
\begin{align}
\label{sup_sn_Lzero}
 dL_0 = \gamma L_0^\alpha d\tau
 \ 
\end{align}
defines the solution of the deterministic dynamics. With the target state $L_\mathrm{ts}=0$ at the boundary, the solution is given as
\begin{equation}
\label{sup_x0_res}
L_0 = \left( (1-\alpha) \gamma \tau \right)^{\frac{1}{1-\alpha}} 
\qquad \mathrm{with} \ \alpha <1
\ .
\end{equation}
The equation for the first correction to the deterministic solution Eq.~\eqref{sup_dx1_mult} reads 
\begin{align}
\label{sup_dx1_expl_mult}
 d L_1 = L_1 k\left(L_0(\tau)\right) d\tau + c(L_0(\tau)) d W_\tau
 \ .
\end{align}
The time dependent drift coefficient evaluates to
\begin{align}
\label{sup_k}
  k\left(L_0\right) = \frac{d a(L_0)}{d L_0} = \gamma \frac{d L_0^\alpha}{d L_0}
	= \gamma \alpha L_0^{\alpha-1}
	= \gamma \alpha \left( (1-\alpha) \gamma \tau \right)^{-1}
\ ,
\end{align}
and the coefficient of the diffusion term is given as
\begin{align}
\label{sup_c_mult}
 c(L_0(\tau) )
 = 
 L_0^\frac{\beta}{2} 
 = 
\left( (1-\alpha) \gamma \tau \right)^{\frac{\beta}{2-2\alpha}} 
\ ,
\end{align}
where for both we used the explicit solution for $L_0$ stated in Eq.~\eqref{sup_x0_res}.
The formal solution to Eq.~\eqref{sup_dx1_expl_mult} then reads
\begin{align}
\label{sup_x1_res_mult}
 L_1(\tau) 
 &= \int_0^\tau  
 c(L_0(\tau'))
 e^{
 \int_{\tau'}^\tau k(L_0(s)) ds
 }
 d W_{\tau'}
 \ ,
\intertext{or explicitly written}
 &= 
 \int_0^\tau
 \left( (1-\alpha) \gamma \tau' \right)^{\frac{\beta}{2-2\alpha}} 
 \left(\frac{\tau}{\tau'}\right)^{-\frac{\alpha
   }{\alpha -1}}
   d W_{\tau'} 
   \ ,
\end{align}
 with $k(L_0(s))$ taken from Eq.~\eqref{sup_k} and $c(L_0(\tau'))$ from Eq.~\eqref{sup_c_mult}. As above, we assume the target state at $L(0)=0$ and thus $L_1(0)=0$.

To formally find an analytic expression for the second order contribution $L_2(\tau)$ we must solve Eq.~\eqref{sup_dx2_mult}. However, as we will see below, to construct expressions for mean, variance and two-time covariance up to order $D$ it is sufficient to determine the ensemble average of $L_2(\tau)$. Averaging over Eq.~\eqref{sup_dx2_mult}, we arrive at the ordinary differential equation (ODE)
\begin{equation}
\label{eq:l2_ode}
 \frac{d \langle L_2 \rangle}{d\tau}
 =
 \frac{d a(L_0)}{dL_0} \langle L_2 \rangle
 + \frac{1}{2} \frac{d^2 a(L_0)}{d L_0^2} \langle L_1^2 \rangle
 + b(L_0)
 \ ,
\end{equation}
where $b(L_0)$ is given as
\begin{align}
 b(L_0) = - (\alpha-\beta) L_0^{\beta-1}
 .
\end{align}
We next exploit the expressions for $L_0$, $L_1$ and $\langle L_2 \rangle$ to construct moments for the homogeneous dynamics. We then use the obtained expressions to generalized the results to heterogeneous force laws.

For the ensemble mean we substitute $L$ with its expansion up to order $D$ to obtain
\begin{align}
\label{sup_def_sn_mean}
\langle L(\tau) \rangle 
&= 
\langle L_0(\tau) \rangle 
+ \sqrt{D} \langle L_1(\tau) \rangle 
+ D \langle L_2(\tau) \rangle
+ \mathcal{O}(D^\frac{3}{2}) \ .
\intertext{The zero-th order term is simply the deterministic solution. The next leading order term 
$\langle L_1(\tau) \rangle$ evaluates to zero, as the Wiener increment denotes a zero mean white noise stochastic process. The mean up to order $D$ is thus constituted as
}
&= 
\langle L_0(\tau) \rangle 
+ D \langle L_2(\tau) \rangle
+ \mathcal{O}(D^\frac{3}{2}) \ .
\end{align}
For the variance 
\begin{align}
\label{sup_def_sn_var}
\sigma_L^2(\tau)
&=
\langle L(\tau)^2 \rangle - \langle L(\tau) \rangle^2
=
\langle \left(L_0(\tau) + \sqrt{D}L_1(\tau) + D L_2(\tau) \right)^2 \rangle 
- 
\langle L_0(\tau) + \sqrt{D}L_1(\tau) + D L_2(\tau) \rangle^2
+ \mathcal{O}(D^\frac{3}{2})
\intertext{
only one term up to order $D$ survives. 
The zero-th order term is a deterministic expression and thus evaluates to zero. Furthermore, the $L_0$ $L_1$ cross terms evaluate to zero with $L_0$ deterministic and 
$\langle L_1(\tau) \rangle =0$. Analogously with $L_0$ deterministic the cross $L_0$ $L_2$ term cancels. The only surviving term is thus the order $D$ term
}
&=
D \langle L_1^2(\tau) \rangle + \mathcal{O}(D^\frac{3}{2})
\end{align}
For the two-time covariance the same observations as for the variance hold. Only the doubly $L_1$ dependent term
\begin{align}
  C(\tau,\tau') &= 
\langle
\left(
L(\tau) - \langle L(\tau) \rangle
\right)
\left(
L(\tau') - \langle L(\tau') \rangle 
\right)
\rangle
=
D \langle L_1(\tau) L_1(s) \rangle + \mathcal{O}(D^\frac{3}{2})
\end{align}
survives. 
We next evaluate all contributing expressions for mean, variance and two-time covariance. The combined expressions are stated in the main text. For the sole contribution to the variance we find
\begin{equation}
\label{sup_var_epsilon_expansion_mult}
 \langle L_1^2(\tau) \rangle 
 =  
\int_0^\tau
 \left( (1-\alpha) \gamma \tau' \right)^{\frac{\beta}{1-\alpha}} 
 \left(\frac{\tau}{\tau'}\right)^{-\frac{2 \alpha
   }{\alpha -1}}
   d \tau'
   =
   \frac{1}{\gamma}
   \frac{ ((1 -\alpha)\gamma \tau)^{\frac{1-\alpha+\beta }{1-\alpha }}}{1-3 \alpha
   +\beta}
   \ .
\end{equation}
This expression is strictly valid for $\alpha<\beta-1$ with $\beta<2$. This is in accordance with the range of validity of the expanded Langevin equation (Eq.~\eqref{sup_tr_lv_tsa_sn_multN_alpha_sm_beta_simp}). Using the more loose range of validity for Eq.~\eqref{sup_tr_lv_tsa_sn_multN_alpha_sm_beta_simp} ($\alpha<\beta$), the pole at $1-3\alpha+\beta$ becomes attainable for $\alpha > 0.5$ and the approximation breaks down.    

Similar to the variance, the two time covariance is built from the product of two $L_1(\tau)$ terms. Evaluating these terms we find
\begin{align}
\label{sup_2time_cov_epsilon_expansion_mult}
 \langle L_1(\tau) L_1(\tau') \rangle
 &=
\int_0^{\mathrm{min}(\tau,\tau')} 
 \left( (1-\alpha) \gamma s \right)^{\frac{\beta}{1-\alpha}} 
  \left(\frac{\tau}{s}\right)^{-\frac{ \alpha
   }{\alpha -1}}
     \left(\frac{\tau'}{s}\right)^{-\frac{ \alpha
   }{\alpha -1}}
   d s
\notag
   \\
   &=
\begin{cases}
  \frac{((1-\alpha ) \gamma  \tau')^{-\frac{\alpha }{\alpha -1}} ((1-\alpha ) 
   \gamma \tau)^{\frac{1 -2 \alpha +\beta}{1-\alpha }}}{\gamma \left(1 -3 \alpha +\beta \right)}
   &\qquad \mathrm{for} \ \tau < \tau'
   \\
  \frac{((1-\alpha ) \gamma \tau)^{-\frac{\alpha }{\alpha -1}} ((1-\alpha )  
   \gamma \tau')^{\frac{1 -2 \alpha +\beta}{1-\alpha }}}{\gamma \left(1 -3 \alpha +\beta \right)}
   &\qquad \mathrm{for} \ \tau > \tau' 
   \\
    \frac{( (1 -\alpha) \gamma \tau)^{\frac{1 -\alpha +\beta}{1-\alpha }}}{
 \gamma \left(1 - 3 \alpha + \beta \right) }
   &\qquad \mathrm{for} \ \tau = \tau'
   \ .
\end{cases}
\end{align}
The range of validity is the same as for the variance.

In the last step we solve the ODE Eq.\eqref{eq:l2_ode} to find the order $D$ contribution to the mean 
%
%
%
%
\begin{equation}
\label{eq:sol_l2}
\langle L_2(\tau) \rangle
=
\frac{\left(7 \alpha ^2-\alpha  (8 \beta +3)+2 \beta  (\beta +1)\right) ((1-\alpha) \gamma 
   \tau)^{\frac{\alpha -\beta }{\alpha -1}}}{2 \gamma  (2 \alpha -\beta ) (3 \alpha -\beta
   -1)}
%
\ .
\end{equation}
For $\alpha<\beta-1$ with $\beta<2$ none of the poles can be reached and the expression is valid. For the extended range which only requires $\alpha<\beta$, the pole at $2\alpha -\beta$ can be reached for $\alpha>0$.
The second pole is identical to the one of the variance and discussed there.

The discussion of the poles in the expression for mean, variance and covariance suggests to use the derived expression genuinely for $\alpha<\beta$ but with slightly more care in the region $\beta-1<\alpha<\beta$ if $0<\alpha<1$ holds.

\section{Small noise expansion for heterogeneous dynamics}  
In this section we discuss how heterogeneity quantitatively changes the moments obtained from the small noise expansion of homogeneous dynamics. 
We use the splitting of the total average into a conditional average with fixed $\gamma$ and an average with respect to $P(\gamma)$, as defined in the main text. This allows us to generalize the expansion terms obtained in section \ref{sec_hom_sN_exp}. As in the main text, we denote averages with respect to parameters by $\left[ \cdot \right]$

Using this approach, the expansion of the mean for homogeneous ensembles, as stated in Eq.~\eqref{sup_def_sn_mean}, generalizes to



\begin{align}
\label{moment_expansion_constD_hetDyn_mean}
\overline{L}(\tau)
&=
 \left[ \ \langle L(\tau) \rangle \ \right]
 =
 \left[  L_0(\tau)  \right]
 +
 D \left[ \ \langle L_2(\tau) \rangle \ \right]
 + 
 \mathcal{O}(D^\frac{3}{2})
\ .
\end{align}
As for the homogeneous case, the term of order 1, that is $\langle L_1(\tau)\rangle$, does not appear in Eq.~\eqref{moment_expansion_constD_hetDyn_mean}.
It denotes the average over a zero mean stochastic process and thus evaluates to zero already on the subsample level. The two remaining contributions can directly be substituted into Eq.~\eqref{moment_expansion_constD_hetDyn_mean} using the results stated in Eq.~\eqref{sup_x0_res} and Eq.~\eqref{eq:sol_l2} for the homogeneous case. The resulting averaged expression is stated in the main text.

Compared to the homogeneous case, the variance
\begin{align}
\label{moment_expansion_constD_hetDyn_var}
\sigma^2_L(\tau)
&=
 \left[ \ \langle 
 L(\tau)^2  
\rangle \ \right]
-
 \left[ \ \langle 
 L(\tau) 
\rangle \ \right]^2
 \intertext{
 does now also include relevant contributions of 0th order, that is $L_0$-dependent terms. Terms linear in $\langle L_1(\tau) \rangle$ still evaluate to zero. We find
}
 &=
 \left[ 
 L_0(\tau)^2  
 \right]
-
 \left[  
 L_0(\tau) 
\right]^2
\notag\\
 & \quad +
 D \left[ \ \langle L_1(\tau)^2 \rangle \ \right]
\notag\\ 
 & \quad +
 2 D \left(
 \
    \left[ \  
    L_0(\tau)
    \langle
        L_2(\tau) 
    \rangle \ \right]
     - 
     \left[
        L_0(\tau)
     \right] 
    \left[ \ \langle
        L_2(\tau) 
    \rangle \ \right] 
    \
  \right)
 \notag\\ 
 & \quad +
 \mathcal{O}\left(D^\frac{3}{2} \right)
 \ .
\intertext{
In particular the zero-th order term, which is defined as the variance of the deterministic solution, changes the observed ensemble variance. The second additional term is of order $D$ and provides the coupling between heterogeneity and stochasticity. In the final step, we rearrange terms 
}
 &=
 \left[ 
 L_0(\tau)^2  
 \right]
-
 \left[  
 L_0(\tau) 
\right]^2
\notag\\
 & \quad +
 D \left[ \ \langle L_1(\tau)^2 \rangle \ \right]
 \Bigg(
%
 1 
 +
  \left(
 \
1
     - 
     \frac{
     \left[
        L_0(\tau)
     \right] 
    \left[ \ \langle
        L_2(\tau) 
    \rangle \ \right] }
    {
        \left[ \  
    L_0(\tau)
    \langle
        L_2(\tau) 
    \rangle \ \right]
    }
    \
  \right)
  \frac{
     2 \left[ \  
    L_0(\tau)
    \langle
        L_2(\tau) 
    \rangle \ \right]}{
    \left[ \ \langle L_1(\tau)^2 \rangle \ \right]
    }
  \Bigg)
 \notag\\ 
 & \quad +
 \mathcal{O}\left(D^\frac{3}{2} \right)
 \ ,
\end{align}
to end up with the form used in the main text.

For the covariance 
\begin{align}
\label{moment_expansion_constD_hetDyn_cov}
C(\tau,\tau')
&=
 \left[ \ \langle 
 \ \left( \
 L(\tau)  
 -
 \left[ \ \langle L(\tau) \rangle \ \right]
 \ \right) 
 \ \left( \
 L(\tau')  
 -
 \left[ \ \langle L(\tau') \rangle \ \right]
 \ \right) \ 
 \rangle \ \right]
\intertext{the same considerations and couplings between the expansion variables apply as for the variance. We find}
%
 &=
  \left[
  \
 \left( \
 L_0(\tau)  
 -
 \left[  L_0(\tau)  \right]
 \ \right) 
 \ \left( \
 L_0(\tau')  
 -
 \left[  L_0(\tau')  \right]
 \ \right) 
 \
 \right]
 \notag \\
  & \quad + 
D
 \left[ \ \langle 
 L_1(\tau) 
L_1(\tau')
 \rangle \ \right]
 \notag \\
  & \quad + 
  D
   \left( \
    \left[ \ 
        L_0(\tau)
        \langle
        L_2(\tau') 
    \rangle \ \right]
     - 
     \left[
        L_0(\tau)
    \right] 
    \left[ \ \langle
        L_2(\tau') 
    \rangle \ \right] 
  \ \right)
 \notag \\
  & \quad + 
  D
   \left( \
    \left[ \ 
        L_0(\tau')
        \langle
        L_2(\tau) 
    \rangle \ \right]
     - 
     \left[ \ 
        L_0(\tau')
    \right] 
    \left[ \ \langle
        L_2(\tau) 
    \rangle \ \right] 
    \
  \ \right)
  + 
 \mathcal{O}(D^\frac{3}{2})
\ .
\intertext{
Compared to the homogeneous case, the covariance of the heterogeneous deterministic solutions is added, and the covariance coupling between heterogeneity and stochasticity. Rearranging terms}
     &=
  \left[
  \
 \left( \
 L_0(\tau)  
 -
 \left[  L_0(\tau)  \right]
 \ \right) 
 \ \left( \
 L_0(\tau')  
 -
 \left[  L_0(\tau')  \right]
 \ \right) 
 \
 \right]
 \notag \\
  & \quad + 
D
 \left[ \ \langle 
 L_1(\tau) 
L_1(\tau')
 \rangle \ \right]
 \notag \\
& \quad + 
D \left[ \ \langle L_1(\tau)^2 \rangle \ \right]
  \left(
 \
1
     - 
     \frac{
     \left[
        L_0(\tau)
     \right] 
    \left[ \ \langle
        L_2(\tau') 
    \rangle \ \right] }
    {
        \left[ \  
    L_0(\tau)
    \langle
        L_2(\tau') 
    \rangle \ \right]
    }
    \
  \right)
  \frac{
      \left[ \  
    L_0(\tau)
    \langle
        L_2(\tau') 
    \rangle \ \right]}{
    \left[ \ \langle L_1(\tau)^2 \rangle \ \right]
    }
  \notag \\ 
& \quad +
   D \left[ \ \langle L_1(\tau')^2 \rangle \ \right]
  \left(
 \
1
     - 
     \frac{
     \left[
        L_0(\tau')
     \right] 
    \left[ \ \langle
        L_2(\tau) 
    \rangle \ \right] }
    {
        \left[ \  
    L_0(\tau')
    \langle
        L_2(\tau) 
    \rangle \ \right]
    }
    \
  \right)
  \frac{
      \left[ \  
    L_0(\tau')
    \langle
        L_2(\tau) 
    \rangle \ \right]}{
    \left[ \ \langle L_1(\tau')^2 \rangle \ \right]
    }
  + 
 \mathcal{O}(D^\frac{3}{2})
\ ,
\end{align}
we end up with the version used in the main text.
With
$L_0(\tau)$, $\langle L_1^2(\tau) \rangle$, $\langle L_1(\tau)L_1(\tau') \rangle$  and  $\langle L_2(\tau) \rangle$ already derived for the homogeneous case and stated in Eq.~\eqref{sup_x0_res}, Eq.~\eqref{sup_var_epsilon_expansion_mult}, 
Eq.~\eqref{sup_2time_cov_epsilon_expansion_mult} and Eq.~\eqref{eq:sol_l2}, the heterogeneous moments can be calculated. We state the result in the main text.

\section{Covariance for noise dominated TSA dynamics}
In this section we provide the relevant distributions to calculate the covariance for the noise dominated case. For convenience we start with a reminder of the Langevin equation 
 \begin{align}
\label{sup_NoiseDominatedLangevin}
  dL(\tau) =
      D L^{\beta-1}
       d\tau
+
      \sqrt{D \, L^\beta} \ d W_\tau
      \qquad (\beta <2)
\end{align}
for which we want to obtain the covariance.

The covariance for the noise dominated case can be obtained from the joit-probability distribution
\begin{align}
R^N(x_2,\tau_2; x_1,\tau_1)
=
R^N(x_2,\tau_2| x_1,\tau_1) \ R^N(x_1,\tau_1).
\end{align}
For readability, and to avoid confusion between coordinates and expansion terms, we chose to use $x_1$ and $x_2$ to denote the spatial coordinate $L$ at the respective times $\tau_1$ and $\tau_2$ with $\tau_2 > \tau_1$.
We obtained both the transition probability density $R^N(x_2,\tau_2| x_1,\tau_1)$ and the density $R^N(x_1,\tau_1)$ from a coordinate transformation of the respective distributions for the Bessel processes as stated in \cite{bray2000random}. The details of the transformation are stated in our previous text \cite{lenner2023reverse2}. We found for the joint probability density
\begin{align}
\label{sip_joint_dist_noiseDom}
&R(x_2,\tau_2; x_1,\tau_1)
=
\notag \\
&\quad =
\frac{2^{\frac{1}{2-\beta }+2} x_2^{3/2} x_1^{\beta -\frac{3}{2}} (x_1 x_2)^{-\beta }
   \left((\beta -2)^2 D \tau_1 x_1^{\beta -2}\right)^{\frac{1}{\beta -2}-1} \exp \left(- \frac{2
   (x_1 x_2)^{-\beta } \left(\tau_1 x_2^2 x_1^{\beta }+\tau_2x_1^2 x_2^{\beta }\right)}{(\beta -2)^2 D \tau_1
   (\tau_2-\tau_1)}\right) I_{\frac{1}{2-\beta }}\left(\frac{4 x_1 x_2 (x_1 x_2)^{-\beta /2}}{D (\tau_2-\tau_1)
   (\beta -2)^2}\right)}{D  \ \Gamma \left(\frac{\beta -3}{\beta -2}\right) (\tau_2-\tau_1)}
   \ .
\end{align}
Here, $I_{n}\left(z \right)$ denotes the modified Bessel function of the first kind and  $\Gamma\left(n\right)$ the gamma function.

The covariance can now be calculated from the joint distribution Eq.~\eqref{sip_joint_dist_noiseDom}. As stated in the main text, we did not find a closed-form solution for general $\beta$. For a fixed value of $\beta$ with $\beta<2$, results are however obtainable. We state the case $\beta=1$ in the main text and show cuts through the covariance for $\beta=0$ and $\beta=-1$. The covariance for $\beta=0$ is given as
\begin{align}
C_L^{(D)}(\tau_2,\tau_1)
=
\frac{2 D \left(-4 \sqrt{\tau_1 \tau_2}+3 \sqrt{\tau_1 (\tau_2-\tau_1)}+(2 \tau_1+\tau_2) \sin
   ^{-1}\left(\sqrt{\frac{\tau_1}{\tau_2}}\right)\right)}{\pi }
\end{align}
where we only consider the case $\tau_2 > \tau_1$. The other two cases follow directly. For $\tau_2=\tau_1$ the covariance becomes the variance, and for $\tau_2 > \tau_1$ the same covariance expression holds with $\tau_1$ and $\tau_2$ switched.

The covariance for $\beta=-1$ is a bit more unwieldy an reads  
\begin{align}
\label{cov_betaMinusOne}
C_L^{(D)}(\tau_2,\tau_1)
=
D^{2/3}
\frac{\frac{9  \Gamma \left(\frac{5}{6}\right)^2 (\tau_2-\tau_1) \left(4 \tau_2 \left(2
   \tau_1^2+\tau_2^2\right) \, _2F_1\left(\frac{1}{3},\frac{5}{6};\frac{1}{6};\frac{\tau_1^2}{(\tau_1-2
   \tau_2)^2}\right)-(\tau_1-2 \tau_2)^2 (\tau_1+\tau_2) \,
   _2F_1\left(-\frac{2}{3},\frac{5}{6};\frac{1}{6};\frac{\tau_1^2}{(\tau_1-2
   \tau_2)^2}\right)\right)}{\pi  (-\tau_1 (\tau_1-2 \tau_2))^{5/3}}-\frac{4 \sqrt[3]{2} \pi  \sqrt[3]{ \tau_1
   \tau_2}}{\Gamma \left(\frac{7}{6}\right)^2}}{2\cdot 6^{2/3}}
   %
\end{align}
for the case with $\tau_2 > \tau_1$. The term $_2F_1\left(a,b;c;z\right)$ denotes the hypergeometric function.
The case $\tau_1 > \tau_2$ is simply obtained by switching $\tau_1$ for $\tau_2$ and vice versa in expression Eq.~\eqref{cov_betaMinusOne}. For $\tau_1=\tau_2$ we obtain the variance, which in its general form is stated in the main text.


\bibliography{literature_SI}

\newpage

\appendix
\section{Covariance examples from simulations and theory}
In this section we present supplementary figures of results for the covariance, comparing our analytical calculations with simulations of the forward dynamics after target state alignment.

\begin{figure}[ht]
 \centerline{\includegraphics[width=0.6\linewidth]
 {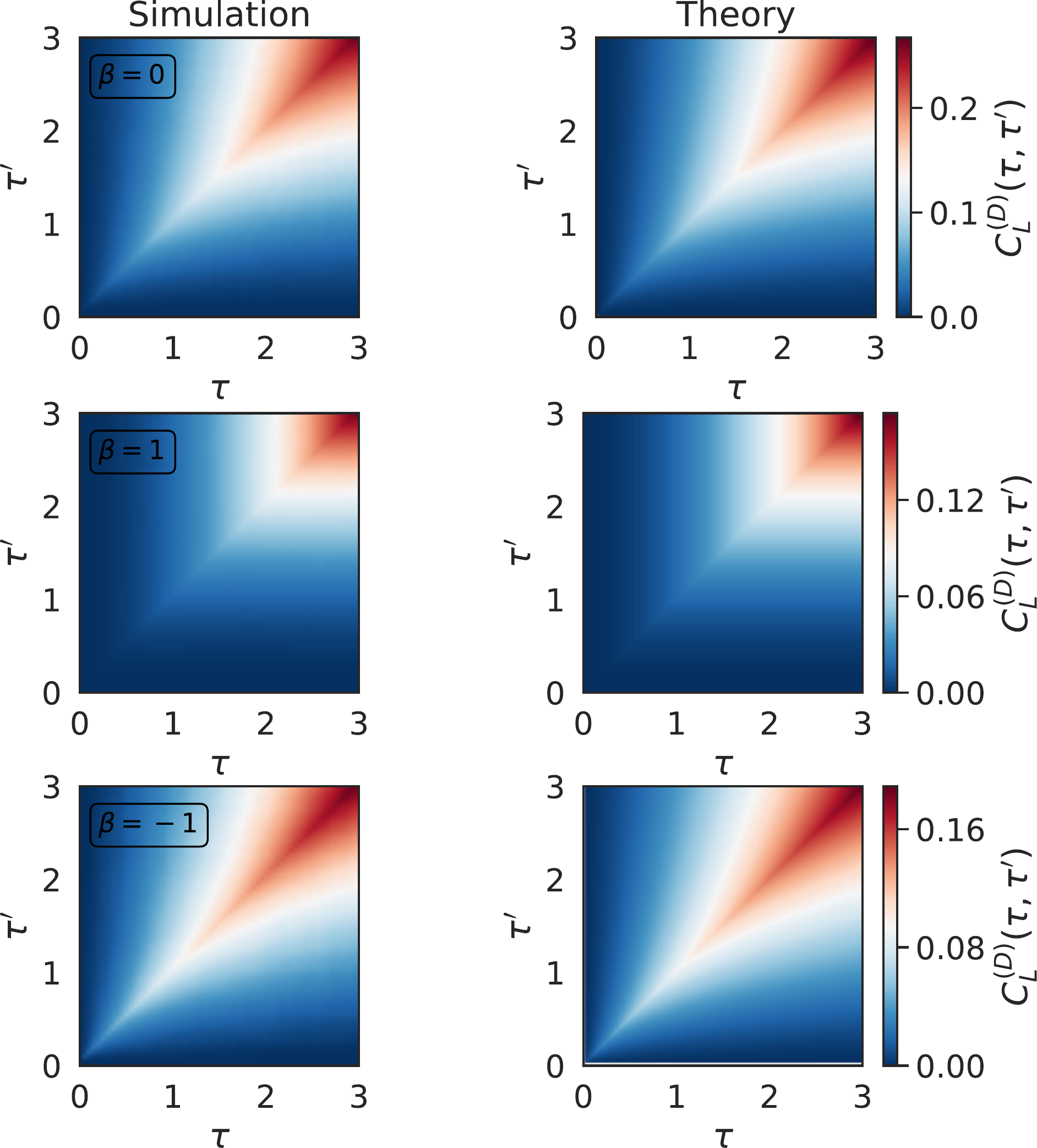}}
 \caption{
\textbf{Covariance for homogeneous noise dominated TSA dynamics.}
 Shown are covariance matrices for dynamics with identical force law $f(L)=-\gamma L^\alpha$ (here $\alpha=-1$),
 but different multiplicative noise $D(L)=D L^\beta$ with $\beta=0,1,-1$.
The ensemble statistics of the forward dynamics has been simulated with 20000 trajectories starting at $\widehat{L}_0=3$. The analytic expressions can be found in the main text. The parameters are $\gamma=1$ and $D=0.2$.
 }\label{fig:NoiseDrivenTSA}
 \end{figure}

\begin{figure}[ht]
 \centerline{\includegraphics[width=0.6\linewidth]
 {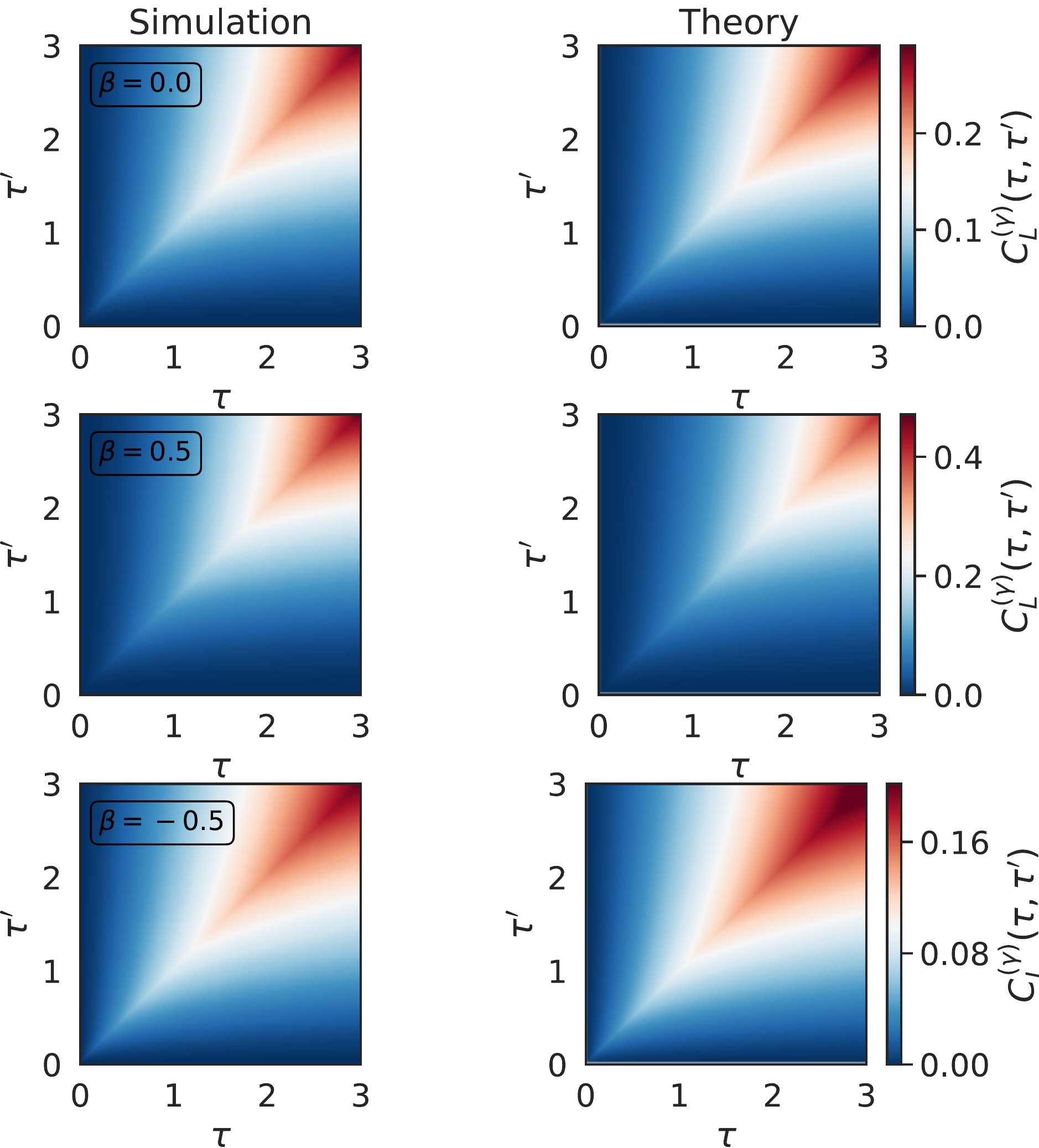}}
 \caption{
\textbf{Covariance for homogeneous force dominated TSA dynamics.} We here show covariance matrices for TSA dynamics with identical force laws $f(L)=-\gamma L^\alpha$ (here $\alpha=-1$),
 but different multiplicative noise $D(L)=D L^\beta$ with $\beta=0,0.5,-0.5$.
 The ensemble statistics of the forward dynamics has been simulated with 20000  trajectories that start at $\widehat{L}_0=20$. The analytic expressions can be found in the main text. Parameters are $\gamma=1$ and $D=0.2$.
 }\label{fig:ForceDrivenTSA}
 \end{figure}

\begin{figure}[ht]
 \centerline{\includegraphics[width=0.6\linewidth]
 {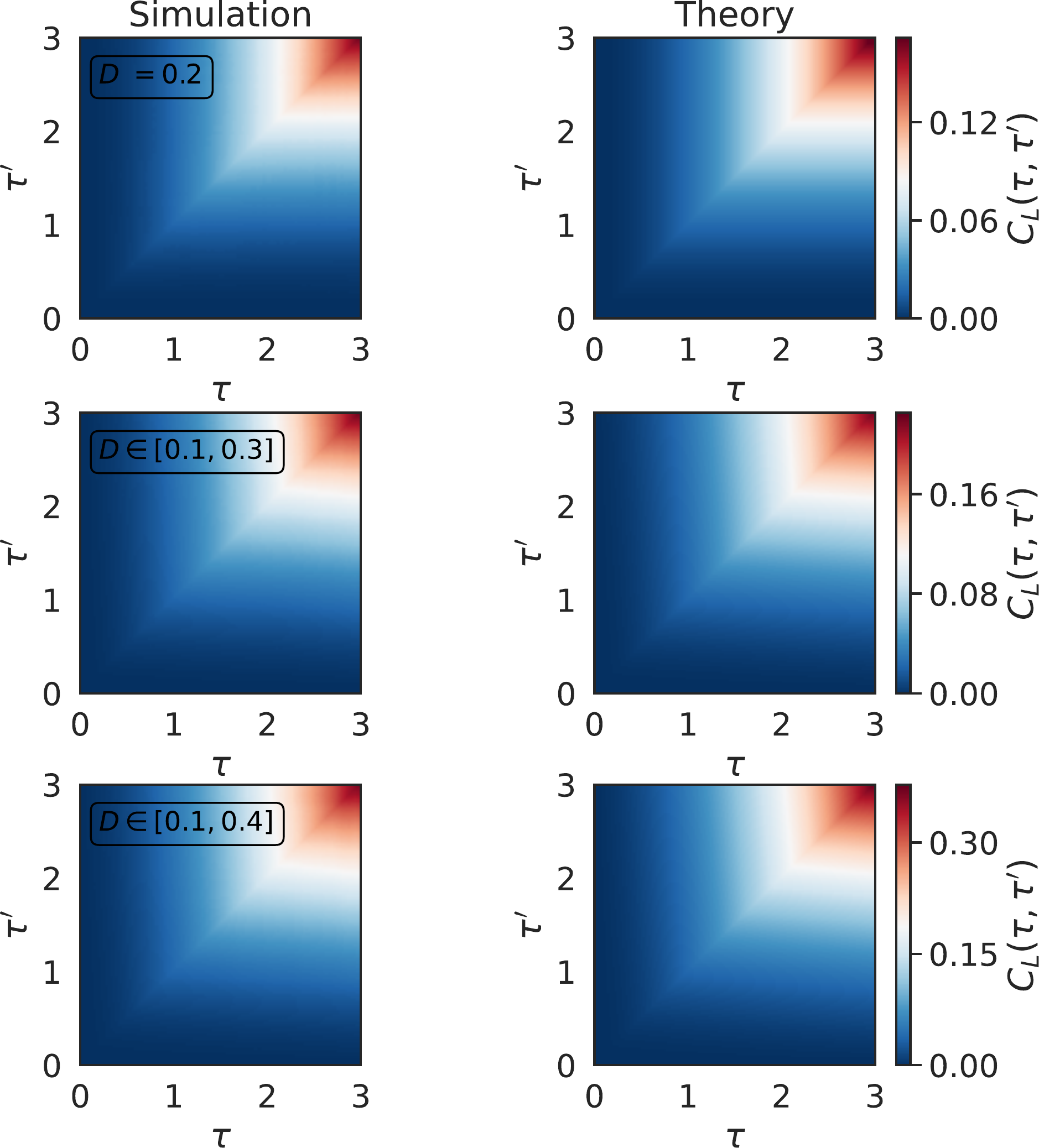}}
 \caption{
\textbf{Covariance for heterogeneity in the diffusion constant $D$ of noise dominated TSA dynamics.}
 We compare the homogeneous case with $D(L)= D L$  studied in Fig.~\ref{fig:NoiseDrivenTSA} ($D=0.2$), to the heterogeneous version with $D$ randomly drawn from a fixed interval for each sample path realization. 
 We chose $D\in \left[0.1 , 0.3 \right]$ and $D\in \left[0.1 , 0.4 \right]$.
The statistics of the forward dynamics are based on 15000 simulated trajectories that start at $\widehat{L}_0=6$. The analytic expressions can be found in the main text. The parameter $\gamma=0$ was chosen here.
 }\label{fig:NoiseDrivenTSA_het}
 \end{figure}

\begin{figure}[ht]
 \centerline{\includegraphics[width=0.6\linewidth]
 {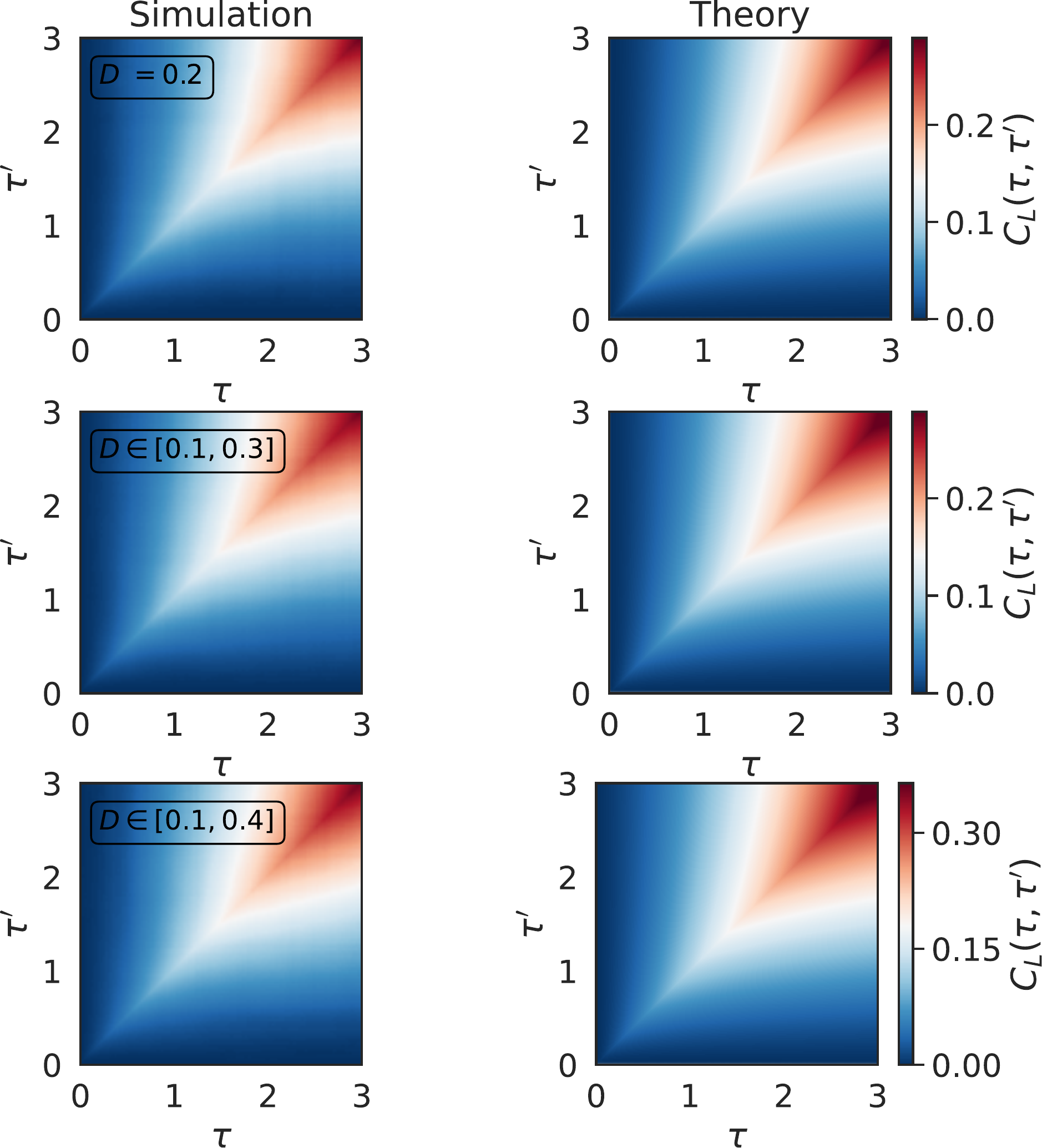}}
 \caption{
\textbf{Covariance for heterogeneity in the noise strength $D$ of force dominated TSA dynamics.}
We compare the ensemble statistics of homogeneous dynamics  with $f(L)=-\frac{\gamma}{L}$, $D(L)= D$, to the heterogeneous case with $D$ randomly drawn from a fixed interval for each sample path realization. Chosen are $D \in [0.1,0.3]$, $D \in [0.1,0.4]$ and for comparison the homogeneous case with $D=0.2$.
 The statistics of the target state aligned forward simulations are based on 20000 trajectories that start at $\widehat{L}_0=20$. The analytic expression for the mean, variance and covariance can be found in the main text. The force strength is chosen as $\gamma=1$.
 }\label{fig:ForceDrivenTSA_hetD}
 \end{figure}

\begin{figure}[ht]
 \centerline{\includegraphics[width=0.6\linewidth]
 {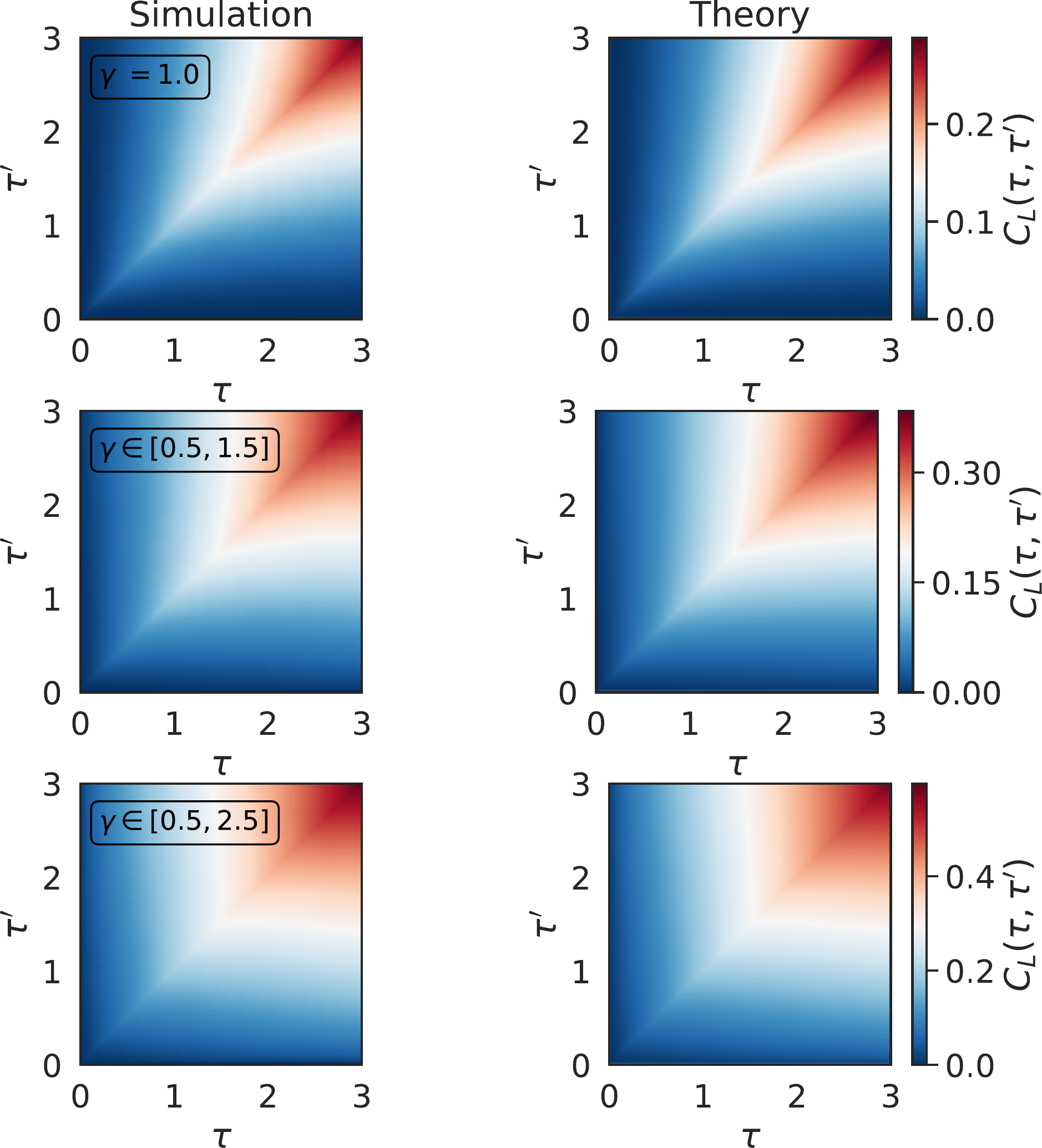}}
 \caption{
\textbf{Covariance for heterogeneity in the force strength $\gamma$ of force dominated TSA dynamics.}
We compare the ensemble statistics of homogeneous dynamics  with $f(L)=-\frac{\gamma}{L}$, $D(L)= D$ studied in Fig.~\ref{fig:ForceDrivenTSA}, to the heterogeneous case with $\gamma$ randomly drawn from a fixed interval for each sample path realization. Chosen are $\gamma \in [0.5,1.5]$, $\gamma \in [0.5,2.5]$ and for comparison the homogeneous case with $\gamma=1.0$.
 The statistics of the target state aligned forward simulations are based on 7000 trajectories that start at $\widehat{L}_0=20$. The analytic expressions can be found in the main text. The diffusion constant is chosen as $D=0.2$.
 }\label{fig:ForceDrivenTSA_het}
 \end{figure}
 